\documentclass[12pt]{article} 
\usepackage{showlabels}
\usepackage{amsmath}
  \usepackage{graphicx}
  \usepackage{color}
  \usepackage{subfigure}
  \usepackage{tikz}
  \usepackage{braket}
  \newlength{\abstractwidth}
 \usepackage[numbers,sort&compress]{natbib}
 \usepackage{amsmath,amssymb}
\usepackage{graphicx}
\usepackage{caption}
\usepackage[export]{adjustbox}
\usepackage{caption}
\usepackage{lipsum}
\usepackage[utf8]{inputenc}
  \usepackage{mciteplus} 

\usepackage[colorlinks = true,
            linkcolor = blue,
            urlcolor  = blue,
            citecolor = blue,
            anchorcolor = blue]{hyperref}

\hypersetup{colorlinks=true,linkcolor=blue}

\usepackage[title]{appendix}
\usepackage{comment}
\usepackage{braket}

\def\XXint#1#2#3{{\setbox0=\hbox{$#1{#2#3}{\int}$}
     \vcenter{\hbox{$#2#3$}}\kern-.5\wd0}}

\makeatletter
\def\blfootnote{\xdef\@thefnmark{}\@footnotetext}
\makeatother

  \newcommand{\be}{\begin{equation}}
  \newcommand{\bea}{\begin{eqnarray}}
  \newcommand{\eea}{\end{eqnarray}}
  \newcommand{\beq}{\begin{equation}}
  \newcommand{\ee}{\end{equation}}
  \newcommand{\eeq}{\end{equation}}

\def\la{\label}

  \def\ba{\begin{eqnarray}}
  \def\ea{\end{eqnarray}}

 \def\simleq{\; \raise0.3ex\hbox{$<$\kern-0.75em
      \raise-1.1ex\hbox{$\sim$}}\; }
 \def\simgeq{\; \raise0.3ex\hbox{$>$\kern-0.75em
      \raise-1.1ex\hbox{$\sim$}}\; }

\def\nref#1{(\ref{#1})}

\def \YC#1{{\color{blue}  YC:#1}}
\def \HL#1{{\color{red}  HL:#1}}

\def\sr#1#2{S\lp #1 \middle| #2 \rp}
	\newcommand{\eqn}[1]{\begin{equation}\begin{split} #1 \end{split}\end{equation}}	
	
	\newcommand{\lp}{\left (}
	\newcommand{\rp}{\right )}

	\newcommand{\inv}{^{-1}}
	\newcommand{\tr}{\text{tr} \,}

	\newcommand{\ev}[1]{\left \langle #1 \right \rangle}

	\usepackage{slashed}

\begin{document}

\begin{titlepage}
  \bigskip

  \bigskip\bigskip

  \bigskip

\begin{center}
{\Large \bf{}}
 \bigskip
{\Large \bf Signatures of global symmetry violation in relative entropies and replica wormholes}
\bigskip
\bigskip
   \bigskip
\bigskip
\end{center}

  \begin{center}

 \bf {Yiming Chen and Henry W. Lin}
  \bigskip \rm
\bigskip
 
 \rm

Jadwin Hall, Princeton University,  Princeton, NJ 08540, USA\\

  \end{center}

 \bigskip\bigskip

\begin{abstract}


It is widely believed that exact global symmetries do not exist in theories that admit quantum black holes.
Here we propose a way to quantify the degree of global symmetry violation in the Hawking radiation of a black hole by using certain relative entropies. While the violations of global symmetry that we consider are non-perturbative effects, they nevertheless give $\mathcal{O}(1)$ contributions to the relative entropy after the Page time.
Furthermore, using ``island'' formulas, these relative entropies can be computed within semi-classical gravity, which we demonstrate with explicit examples.
These formulas give a rather precise operational sense to the statement that a global charge thrown into an old black hole will be lost after a scrambling time. 

The relative entropies considered here may also be computed using a replica trick.
At integer replica index, the global symmetry violating effects manifest themselves as charge flowing through the replica wormhole.

 \end{abstract}
\bigskip \bigskip \bigskip
\blfootnote{ymchen.phys@gmail.com,  hwlin@princeton.edu}

  \end{titlepage}

   \tableofcontents



\section{Introduction}

An old conjecture states that in any theory of quantum gravity with black holes, there are no exact global symmetries \cite{Giddings:1988cx,Kallosh:1995hi,ArkaniHamed:2006dz,banks2011symmetries,Harlow:2018tng,Harlow:2018jwu}. 
A semiclassical treatment of black hole evaporation \cite{Hawking:1974sw} suggests that the Hawking radiation has a thermal spectrum, independent of whatever the global charge of the matter that formed the black hole. After the black hole evaporates completely, any initial global charge is lost (up to fluctuations in the thermal ensemble). One might hope that the charge can be stored in some remnant of the black hole, but such remnants would violate entropy bounds \cite{banks2011symmetries} among other problems \cite{Susskind:1995da}. These arguments are most clear at the final stage of the evaporation, where the black hole is not big enough to store the information about the infalling charge, while they are less clear when the black hole is still large, say at the Page time of the black hole evaporation. 
Said differently, an experimentalist who wishes to check whether symmetry violation has occurred must gather almost all of the Hawking radiation from the black hole. If she is unable to capture the high energy modes at the endpoint of evaporation, or if she is unable to access certain fields (e.g. dark matter fields), she may not be able to conclusively convince herself that charge conservation is violated in nature. 

Since we expect that the global symmetry violation happens throughout the evaporating process, it is therefore natural to ask that whether there are some quantities that can be computed with only some portion of the Hawking radiation, that are able to show explicitly the global symmetry has been violated. (There is a strong analogy here with the Page curve \cite{Page:1993wv}: unitarity implies that the entropy of {\it all} the Hawking radiation from a fully evaporated black hole is zero; Page famously showed that unitarity also has implications for a large subset of the radiation.)
One obvious candidate is to compute correlators which would be zero if the global symmetry is exact. However, if there are no explicit symmetry violating processes in the semiclassical physics, these correlators are expected to have non-perturbatively small values $\mathcal{O}(e^{-S_{\rm BH}})$, where $S_{\rm BH}$ is the Bekenstein-Hawking entropy of the black hole. Computing these quantities will require non-perturbative knowledge beyond Hawking's description. To be clear, we are not claiming that the approximate global symmetry, such as the $B-L$ symmetry in nature, cannot be violated within the semiclassical physics. What we are imagining is the worst case scenario, namely even if the symmetry is preserved semiclassically, non-perturbative gravitational effect will eventually violate it.

Although understanding how to compute such exponentially small violations is a worthy goal for the future, the point of this paper is to point out there exist other quantities which quantify the symmetry violation in the Hawking radiation, that can be computed just with the semiclassical knowledge, and give $\mathcal{O}(1)$ results. Our discussion is motivated by the recent progress in deriving the unitary Page curve \cite{Page:1993wv} for an evaporating black hole within the semiclassical description \cite{Penington:2019npb,Almheiri:2019psf}. In those developments, the key tool is the gravitational fine grained entropy formula, or the quantum extremal surface (QES) prescription \cite{Ryu:2006bv,Hubeny:2007xt,Faulkner:2013ana,Engelhardt:2014gca}. When applying the formula to compute the von Neumann entropy of the Hawking radiation after the Page time, one finds an island in the black hole interior that belongs to the entanglement wedge of the Hawking radiation \cite{Almheiri:2019hni}.  
In this paper, we consider the relative entropies between different density matrices of the Hawking radiation $R$. More precisely, we consider the exact density matrix of the Hawking radiation $\rho_1 = \rho_{\rm exact} (R)$, and the density matrix $\rho_2 = U_R(g) \rho_{\rm exact}(R) U_R^\dagger(g)$ after applying a global symmetry transformation $U_R(g), g \in G$ to the radiation. If we form the black hole from neutral (singlet) matter, then if the global symmetry were exact, we would expect $\rho_2 = \rho_1$ since a reduced density matrix of the system should be invariant under the symmetry transformation. Thus a non-zero relative entropy $S(\rho_2|\rho_1) > 0$ is a way to quantify the degree to which the global symmetry is violated.
Similar to the von Neumann entropy, the relative entropy can be computed with only the knowledge of the semiclassical physics \cite{Jafferis:2015del}, and we find it to be an $\mathcal{O}(1)$ result whenever an island arises. We demonstrate the calculation explicitly using simple examples. The relative entropy that we compute quantifies the amount of global symmetry violation in the Hawking radiation, and we discuss how it depends on various parameters of the problem. As we will see in \S\ref{sec:relentropy}, the relative entropy is zero before the Page time, but becomes non-zero at the Page time and grows as more radiation comes out. 

The island in the QES prescription is closely related to the replica wormhole geometry in the replica trick for deriving the entropy using the gravitational path integral \cite{Lewkowycz:2013nqa,Almheiri:2019qdq,Penington:2019kki}. It is thus natural to ask what roles they play in showing that the global symmetry is violated. In \S\ref{sec:replica} we discuss the replica versions of the relative entropy, and point out that they are non-zero exactly due to charge flowing through the replica wormhole. This resonates with a long history of works 
anticipating that Euclidean wormholes violate global symmetries \cite{Kallosh:1995hi,Abbott:1989jw}. In previous arguments, it was never completely clear whether wormholes should be included or not in a particular calculation; here, unitarity demands them to be included. 
We contrast the global symmetry case with a gauge symmetry, which is allowed because charge cannot flow through the wormhole due to the gauge constraint (Gauss's law).
An exact global symmetry of a holographic theory must therefore be realized as a gauge symmetry in gravity. 

We also consider in \S \ref{increasing} a setup where the bulk relative entropy may be enhanced by creating pairs of particles with opposite charges in the bath, and waiting a scrambling time for the particles to fall into the island. This is related to the Hayden-Preskill \cite{Hayden:2007cs} experiment, as we elaborate on in \S \ref{sec:conclusion} along with other discussions and conclusions.

It is worth noting that a recent paper by Harlow and Shaghoulian \cite{Harlow:2020bee} also connected recent developments in the black hole information problem with the violation of global symmetries, using arguments along the lines of  \cite{Harlow:2018jwu,Harlow:2018tng}. Our result is based on similar assumptions as theirs, but in some sense we take one step further by quantifying and computing the amount that an approximate global symmetry is violated by non-perturbative effects. 

\paragraph{Note added:} as we were finishing the manuscript, we learned that \cite{Hsin:2020mfa} are considering similar issues in an upcoming work.

\section{Quantifying global symmetry violation with relative entropy}\label{sec:relentropy}

\subsection{General argument}\label{sec:general}

We start by reviewing the QES prescription and the relative entropy formula. In the context of an evaporating black hole, the QES prescription as applied to the exact density matrix $\rho_{\rm exact}(R)$ of the Hawking radiation states
\begin{equation}\label{vonNeumann}
    S \left( \rho_{\rm exact}(R) \right) = \textrm{Min} \left\{ \textrm{Ext}_{I} \left[ \frac{A(\partial I)}{4G_{N}}  + S (\rho_{\rm semi} (R \cup I)) \right]   \right\},
\end{equation}
where on the right hand side, one extremizes over the generalized entropy function over possible islands in the region where gravity is dynamical, and choose the minimal one if there are multiple extremums. Note that the right hand side of (\ref{vonNeumann}) is calculable within the semiclassical approximation. The region $R\cup I$ on the right hand side is called the entanglement wedge \cite{Czech:2012bh,Wall:2012uf,Headrick:2014cta} of the region $R$. As first recognized in the context of the AdS/CFT correspondence \cite{Jafferis:2015del}, the validity of (\ref{vonNeumann}) among a family of ``nearby'' states implies an equality between the relative entropies in the exact description and the semiclassical description:
\begin{equation}\label{relformula}
    S\left( \rho_{\rm exact}(R) | \sigma_{\rm exact} (R) \right)  =    S\left( \rho_{\rm semi}(R \cup I) | \sigma_{\rm semi} (R \cup I) \right) ,
\end{equation}
where the relative entropy $S(\rho|\sigma)$ is defined as
\begin{equation}
    S(\rho|\sigma) \equiv \textrm{tr} \left[ \rho (\log \rho - \log \sigma)\right],
\end{equation}
which is a non-negative quantity quantifying how different $\rho$ and $\sigma$ are. 
The relative entropy formula has further subleading corrections, but in this paper we will only be focusing on the leading order result as in (\ref{relformula}).

\begin{figure}[t!]
\begin{center}
\includegraphics[width=12cm]{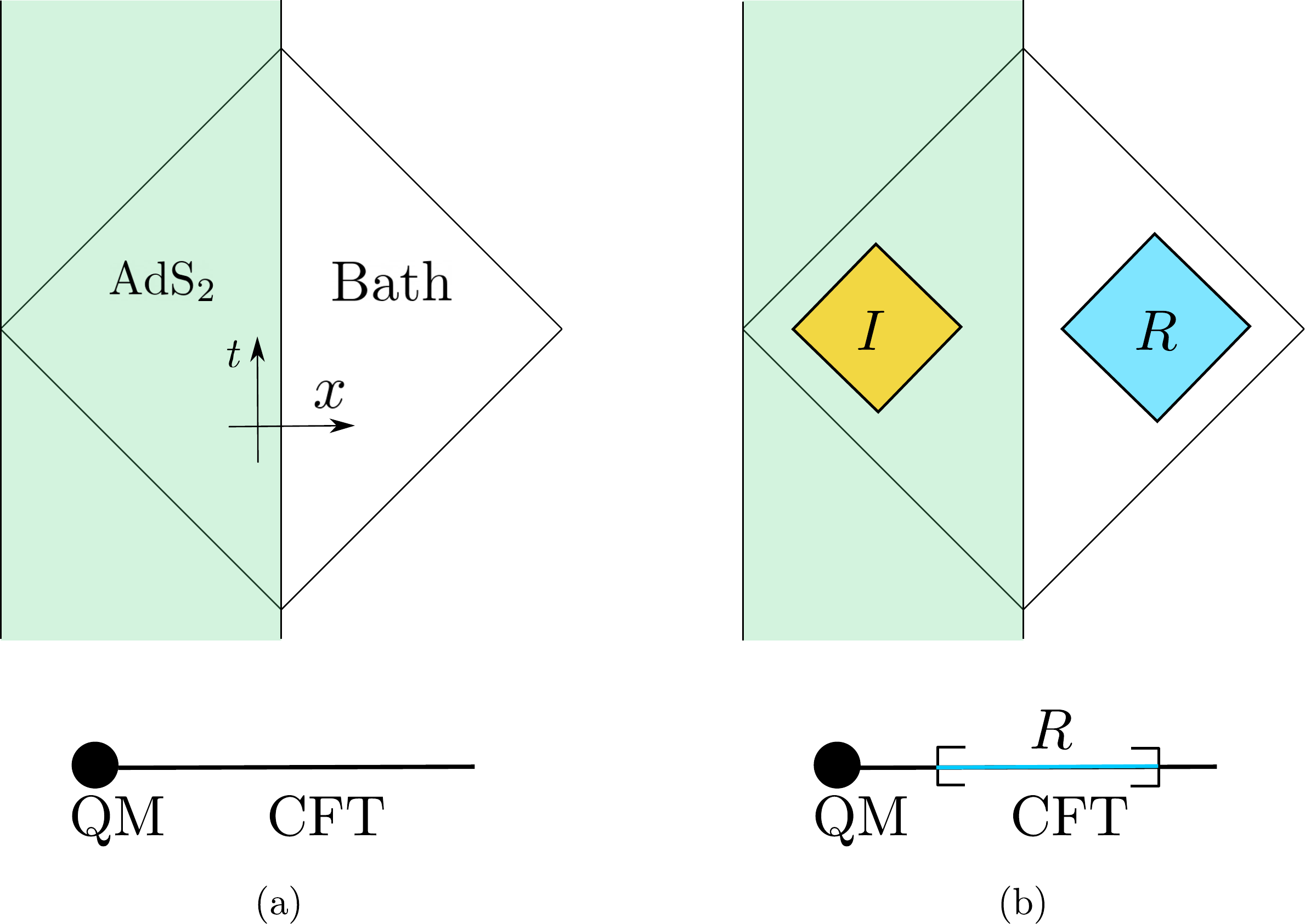}
\caption{(a) A zero temperature black hole in JT gravity coupled to bath. It has a nonperturbative description in terms of a boundary quantum mechanical system coupled to a CFT in half infinite space. (b) When we compute the entropy of a region $R$ in the bath that is large enough, there will be an island $I$ in the gravity region.}
\label{fig:zerotemp}
\end{center}
\end{figure}

Our argument will give non-trivial results in cases where there is an island in the entanglement wedge of the radiation. 
The phenomena of islands has been proposed for a variety of scenarios involving different dimensions and also different signs of the cosmological constant (see \cite{Almheiri:2019yqk,Bousso:2019ykv,Almheiri:2019psy,Akers:2019nfi,Rozali:2019day,Gautason:2020tmk,Anegawa:2020ezn,Hartman:2020swn,Balasubramanian:2020hfs,Hashimoto:2020cas,Geng:2020qvw,Chen:2020uac,Chen:2020tes,Hartman:2020khs,Balasubramanian:2020xqf,Sybesma:2020fxg} for some works in this direction and more in a recent review \cite{Almheiri:2020cfm}). Our argument is general, but we will use the toy examples in \cite{Almheiri:2019yqk} as a concrete set-up for our discussion. More specifically, we will consider the case with a zero temperature black hole in the two-dimensional Jackiw-Teitelboim (JT) gravity \cite{Jackiw:1984je,Teitelboim:1983ux,Almheiri:2014cka} coupled to a flat space bath. There is a two-dimensional CFT with central charge $c \gg 1$ propagating both in the JT gravity region and in the flat space region. This gravity description is semiclassical, namely we will be neglecting nonperturbative effects that are suppressed by $e^{-S_0}$, where $S_0$ is the extremal entropy of the black hole.
We will also assume that the model has a dual nonperturbative description as a quantum mechanical dot coupled to a half infinite line where the CFT lives. The two descriptions of the system are illustrated in Figure \ref{fig:zerotemp} (a). 
For the ground state of the system, the metric and the dilaton profile in the AdS$_2$ region are
\begin{equation}
    ds^2 = \frac{-dt^2 + dx^2}{x^2}, \quad x <0, \quad  \phi = S_0 - \frac{\phi_r}{x},
\end{equation}
where $\phi_r$ defines a length scale of the problem.
The metric in the flat space region is given by
\begin{equation}
    ds^2 = -dt^2 + dx^2, \quad x>0,
\end{equation}
and the conformal fields are in the Minkowski vacuum with respect to the global time coordinate $t$. For a region $R$ in the flat space region that is large enough, applying the QES prescription to compute the entropy (and the location of the entanglement wedge), one finds that the entanglement wedge of $R$ includes an island $I$ in the JT gravity region (see Figure \ref{fig:zerotemp} (b)).

We consider situations in which there is a global symmetry in the semiclassical gravity description, with a symmetry group $G$ that can generally be discrete or continuous, and possibly non-Abelian. We will assume that the symmetry is {\it not} gauged in the gravity description, leaving a discussion of gauge symmetries to \S\ref{sec:gauge}. In terms of the exact ``boundary'' description, the flat space CFT has an exact global symmetry without coupling to the boundary quantum mechanical system, and the coupling breaks this symmetry explicitly. However, the couplings that break the symmetry should be nonperturbatively small so that the global symmetry is still realized in the gravity region in the semiclassical description. 


To simplify the discussion, we will first consider the case where the entire system is in a state which is invariant under the global symmetry transformation in the semiclassical description. For example, we could consider the ground state of the model in Figure \ref{fig:zerotemp} (a), for which the semiclassical state has the global symmetry.\footnote{We will be assuming that the semiclassical global symmetry is not spontaneously broken throughout the paper.} If the state of the entire system is invariant under a symmetry transformation $U(g)$, then when we restrict to a spatial subregion $R$, the reduced density matrix $\rho (R)$ should also be invariant under the symmetry transformation restricted in $R$, which we denote by $U_R(g)$. (We will mostly leave the dependence on the group element $g \in G$ implicit from now on.) The reason is that the symmetry transformation $U$ of a global symmetry has the split property \cite{Harlow:2018tng}; morally, this means it can be written as a product of the symmetry transformation in $R$ and its compliment $\bar{R}$, $U = U_R U_{\bar{R}}$.\footnote{In a continuum quantum field theory, there is an edge term $U_{\rm edge}$ that depends on how we regulate the boundary between $R$ and $\bar{R}$. Since it is not essential to our discussion, we will keep it implicit.} It follows that
\eqn{U_R \rho_R U_R^\dagger &=  \tr_{\bar{R}} \lp  U_R \rho U_R^\dagger \rp =\tr_{\bar{R}} \lp  U_{\bar{R}} U_R \rho U_R^\dagger U_{\bar{R}}^\dagger \rp  
=\tr_{\bar{R}} \lp  U \rho U^\dagger \rp  = \rho_R.}
Applied to our case, since the semiclassical state has the symmetry, we have
\begin{equation}\label{eqn:rhosemiR}
   U_R \rho_{\rm semi} (R) U_{R}^\dagger = \rho_{\rm semi} (R)  ,
\end{equation}
or equivalently
\begin{equation}
     S \left(U_R \rho_{\textrm{semi}} (R)  U_R^\dagger \middle| \rho_{\textrm{semi}} (R) \right)  = 0.
\end{equation}
This motivates us to compare the two density matrices $\rho_{\rm exact} (R)$ and $U_R^\dagger \rho_{\textrm{exact}} (R)  U_R$, by computing the relative entropy between them, i.e. $   S \left( U_R^\dagger \rho_{\textrm{exact}} (R)  U_R\middle| \rho_{\textrm{exact}} (R) \right) $. Interestingly, to compute this quantity, we do not need to have a complete knowledge about $\rho_{\rm exact} (R)$. We can compute it just from the semiclassical description, by applying the relative entropy formula (\ref{relformula}):\footnote{The relative entropy is valid either by considering a transformation $U_R$ that is close to identity in the case of continuous symmetry, or by considering a global symmetry that only acts on a small subset of fields (note that we have $c\gg 1$).}
\begin{equation} \label{relEntropy}
    S \left(  U_R \rho_{\textrm{exact}} (R)  U_R^\dagger \middle| \rho_{\textrm{exact}} (R) \right) =    S \left( U_R \rho_{\rm semi} (R \cup I)  U_R^\dagger \middle|  \rho_{\rm semi} (R \cup I)\right).
\end{equation}

We should stress that the transformation $U_R$ on the left hand side is a completely well defined operator in the nonperturbative description. For a continuous symmetry with Noether current $J_a^\mu$, it is simply given by $e^{i\epsilon_a \int_R * J_a}$. An important assumption in the formula however is that when we pass from the exact description to the semiclassical description in (\ref{relEntropy}), the transformation only acts in the region $R$ but not the island, and still acts as the same symmetry transformation in the semiclassical theory.\footnote{We thank Ahmed Almheiri and Edgar Shaghoulian for asking about this assumption.} One should be cautious about this, because the entanglement wedge of $R$ includes both the region itself and the island, and in principle a general operation in $R$ in the exact description can act both in the region $R$ and the island $I$. Indeed, as advocated in \cite{Chen:2019iro}, a modular flow using the exact modular Hamiltonian of $R$ corresponds to a modular flow on $R \cup I$ with the semiclassical modular Hamiltonian. However, in contrast with the exact modular Hamiltonian, which is expected to be non-local and encodes the fine grained information of the state, the operator $U_R$ factorizes into local pieces and does not contain information about the state. In other words, it is a ``simple'' operation. A more careful argument is the following. We can divide the region $R$ into smaller pieces $\{R_i\}$, and the operator $U_R$ factorizes as
\begin{equation}
    U_R = \prod_{i} U_{R_i}.
\end{equation}
If we divide $R$ such that $R_i$ are small enough, the entanglement wedge of each $R_i$ will only contain the region itself. Thus each $U_{R_i}$ should only act within $R_i$ without any ambiguity, so the product should only act in the union of $R_i$, which is $R$. This is part of the idea that has been used in \cite{Harlow:2018jwu,Harlow:2018tng,Harlow:2020bee} to argue that there cannot be an exact global symmetry. In \S\ref{sec:replica}, we will also justify the formula (\ref{relEntropy}) from the replica computation point of view.

After justifying the formula (\ref{relEntropy}), we now turn to its implications. The density matrix $\rho_{\rm semi} (R\cup I)$ is invariant under the semiclassical global symmetry transformation on the union of $R$ and $I$, but it is \emph{not} invariant under $U_R$ alone. This is because the fields in $R$ are entangled with the fields in $I$ and so $\rho_{\rm semi} (R\cup I) \neq \rho_{\rm semi} (R) \otimes \rho_{\rm semi} (I)$; otherwise it would be invariant under $U_R$ because of (\ref{eqn:rhosemiR}). For this reason, for a fixed $U_R$, one expects the result to be larger if the region $R$ and the island have larger mutual information. To summarize, we expect that the bulk formula (\ref{relEntropy}) gives a nonzero relative entropy when there is an island: 
\begin{equation}\label{result}
    S \left(  U_R \rho_{\textrm{exact}} (R)  U_R^\dagger \middle| \rho_{\textrm{exact}} (R) \right) =    S \left( U_R \rho_{\rm semi} (R \cup I)  U_R^\dagger \middle|  \rho_{\rm semi} (R \cup I)\right) = \mathcal{O}(1),
\end{equation}
where $\mathcal{O}(1)$ denotes that this quantity is not suppressed by $e^{-S_0}$. As a side note, assuming that the global semiclassical state is invariant under the symmetry, $\rho_{\rm semi}(R\cup I)$ is invariant under $U_R U_I$, so we can also compute the relative entropy via
\begin{equation} \label{relEntropy2}
    S \left(  U_R \rho_{\textrm{exact}} (R)  U_R^\dagger \middle| \rho_{\textrm{exact}} (R) \right) =    S \left( U_I^\dagger \rho_{\rm semi} (R \cup I)  U_I \middle|  \rho_{\rm semi} (R \cup I)\right).
\end{equation}

The quantity $ S \left( U_R \rho_{\rm semi} (R \cup I)  U_R^\dagger \middle|  \rho_{\rm semi} (R \cup I)\right)$ is computable solely using the semiclassical description. Since the von Neumann entropies of $U_R \rho_{\rm semi} (R \cup I)  U_R^\dagger$ and $\rho_{\rm semi} (R \cup I)$ are the same as they only differ by an unitary, the relative entropy can be expressed as the difference of the expectation value of the modular Hamiltonian $K_{\rm semi}(R\cup I) \equiv -\log \rho_{\rm semi} (R\cup I)$:
\begin{equation}\label{DiffModular}
      S \left( U_R \rho_{\rm semi} (R \cup I)  U_R^\dagger \middle|  \rho_{\rm semi} (R \cup I)\right) = \Delta \langle K_{\rm semi}(R\cup I)   \rangle,
\end{equation}
where we take the expectation value of the modular Hamiltonian in the state after acting with $U_R$, and subtract that in the original state. In simple cases, this computation can be carried out explicitly, as we will demonstrate in \S\ref{sec:example}. 

Before we move on to concrete examples, let us offer some qualitative comments. For the relative entropy to be nonzero, we need the region $R$ to be large enough so that $R$ claims the island. If we divide $R$ into small pieces, then when we only look at the relative entropies for the small spatial pieces, we will not see such an effect. 

Similar statement holds if we divide the fields in $R$ into small subsets. For example, imagine the CFT is $N$ copies of free fermions ($N\gg 1$), which has an $U(N)$ global symmetry. We could consider a symmetry transformation $U_R$ that only acts among a small subset of the fermions, say $M$ fermions with $M \sim \mathcal{O}(1)$. If we only look at the reduced density matrix of the $M$ fermions while tracing out the rest, since it does not claim the island, we would conclude that it is invariant under the symmetry transformation. However, the density matrix of all the $N$ fermions will not be invariant by (\ref{result}). This suggests that the main contribution to the order one result in (\ref{result}) comes from correlations between the $M$ fermions and the rest of the fermions. For this to be true, there must be sufficient correlation between the $M$ fermions and the rest of the fields at the non-perturbative level, despite the fact that they are decoupled and uncorrelated in the semiclassical description.\footnote{Here we are neglecting the universal coupling through the boundary graviton in the semiclassical description. In the current model, this can be justified in the limit of $c\gg 1$ and $S_0/c$ not too large.} This correlation can also be quantified using the gravitational fine grained entropy formula, and we comment more on this in Appendix \ref{app:subset}.

\subsection{Example: $U(1)$ global symmetry of Dirac fermions}\label{sec:example}

As a concrete, calculable example, we consider a setup where the CFT is a tensor product of a massless Dirac fermion and some other CFT with central charge $c-1$ which we will not specify. For simplicity, we will assume that in the semiclassical description, this fermion field does not couple to the rest of the CFT directly, and we also neglect the Schwarzian fluctuations. In the semiclassical description, the model has a $U(1)$ global symmetry which rotates the fermion as $\psi\rightarrow e^{i\theta} \psi$. The time component of the conserved current corresponding to this symmetry is the fermion number density:
\begin{equation}
    J_0 \equiv \psi_+^\dagger \psi_+ + \psi_-^\dagger \psi_-,
\end{equation}
and the transformation we apply in $R$ is
\begin{equation}
     U_R \equiv \exp \left( i \alpha Q_R \right) = \exp \left( i \alpha \int_R dx\, J_0\right). 
\end{equation}

As we explained in (\ref{DiffModular}), the relative entropy is given by the change of the expectation value of the semiclassical modular Hamiltonian $K_{\rm semi} (R \cup I)$:
\begin{equation}
\begin{aligned}
      S \left(  U_R \rho_{\textrm{exact}} (R)  U_R^\dagger \middle| \rho_{\textrm{exact}} (R) \right)   & =  \bra{0} U_R^\dagger  K_{\rm semi}(R \cup I) U_R \ket{0} - \bra{0} K_{\rm semi} (R\cup I) \ket{0} ,
\end{aligned}
\end{equation}
where the $\ket{0}$ state is the vacuum state of the fermion in the semiclassical description. Since the fermion is massless, we can neglect the warp factor inside the AdS region. In other words, up to local terms at the boundary of the island that depends on the warp factor,
the modular Hamiltonian $K_{\rm semi}(R \cup I)$ is just the modular Hamiltonian for two disjoint intervals in flat space, whose form was given explicitly in \cite{Casini:2009vk} which we review in appendix. \ref{app:modular}. We will not need the detailed form of the modular Hamiltonian here to understand the calculation, but only need its basic structure. For the vacuum state, the modular Hamiltonian is quadratic in the fermion operators. It can be separated into two pieces, the local piece $K_{\rm loc}$ which couples operators within $R$ or $I$, and the non-local piece $K_{\rm noloc}$ that couples a fermion operator in $R$ to another in $I$. It is not hard to see that 
\begin{equation}
     U_R^\dagger K_{\rm loc}  U_R = K_{\rm loc} ,
\end{equation}
as $K_{\rm loc}$ can be expressed as an integral of the local energy density operator $T(x)$, which is invariant under $U_R$ (the piece in the island is invariant trivially because it is spacelike separated from $U_R$). 
On the other hand, $K_{\rm noloc}$ transforms non-trivially under $U_R$. $K_{\rm noloc}$ has the form
\begin{equation}
    K_{\rm noloc,\pm} = i \int_I dx \, \int_R dy\, K(x,y) (\psi_{\pm,I}^\dagger (x) \psi_{\pm, R} (y) -  \psi_{\pm,R}^\dagger (y) \psi_{\pm, I}(x) ) ,
\end{equation}
where the kernel $K(x,y)$ is given explicitly in Appendix \ref{app:modular}. The symmetries act as
\begin{equation}
   U_R^\dagger \psi_{\pm,I}^\dagger (x) \psi_{\pm, R} (y) U_R = e^{i\alpha}\psi_{\pm,I}^\dagger (x) \psi_{\pm, R}(y) ,\quad  U_R^\dagger \psi_{\pm,R}^\dagger (y) \psi_{\pm, I} (x) U_R = e^{-i\alpha}\psi_{\pm,R}^\dagger (y) \psi_{\pm, I}(x).
\end{equation}
By noting that $\bra{0}\psi_{\pm,I}^\dagger (x) \psi_{\pm, R}(y)\ket{0} = - \bra{0}\psi_{\pm,R}^\dagger (y) \psi_{\pm, I}(x)\ket{0}$, we obtain
\begin{equation}\label{relModular}
\begin{aligned}
      S \left(  U_R \rho_{\textrm{exact}} (R)  U_R^\dagger \middle| \rho_{\textrm{exact}} (R) \right) = (\cos \alpha - 1)  \bra{0} K_{\rm noloc} \ket{0}.
\end{aligned}
\end{equation}
The expression is periodic in $2\pi$ since the transformation with $\alpha/2\pi = n \in Z$ acts trivially. The expectation value of $K_{\rm noloc}$ is computed explicitly in Appendix \ref{app:modular}. In Poincaré coordinates $x$, if we parametrize the regions as $ I = (a_1,b_1), R = (a_2,b_2)$, then
\begin{equation}\label{relresult}
      S \left(  U_R \rho_{\textrm{exact}} (R)  U_R^\dagger \middle| \rho_{\textrm{exact}} (R) \right) =  \sin^2 \frac{\alpha}{2} \left( 1 + \frac{ (2\eta - 1)\arctan \sqrt{\frac{\eta}{1-\eta}} }{ \sqrt{\eta (1-\eta)} } \right), \quad \eta \equiv \frac{(b_1 - a_1) (b_2 -a_2)}{ (a_2 - a_1 ) (b_2 - b_1 )}.
\end{equation}
We see that (\ref{relresult}) diverges as $1/\sqrt{1-\eta}$ when $\eta \rightarrow 1$, where the island and the bath region get closer. This is consistent with the intuition that the result should be larger if there is more mutual information between $R$ and $I$.\footnote{However, it is interesting to note that it grows faster than the mutual information as $\eta\rightarrow 1$, since the mutual information only grows logarithmically.} 

\begin{figure}[t!]
\begin{center}
\includegraphics[width=13cm]{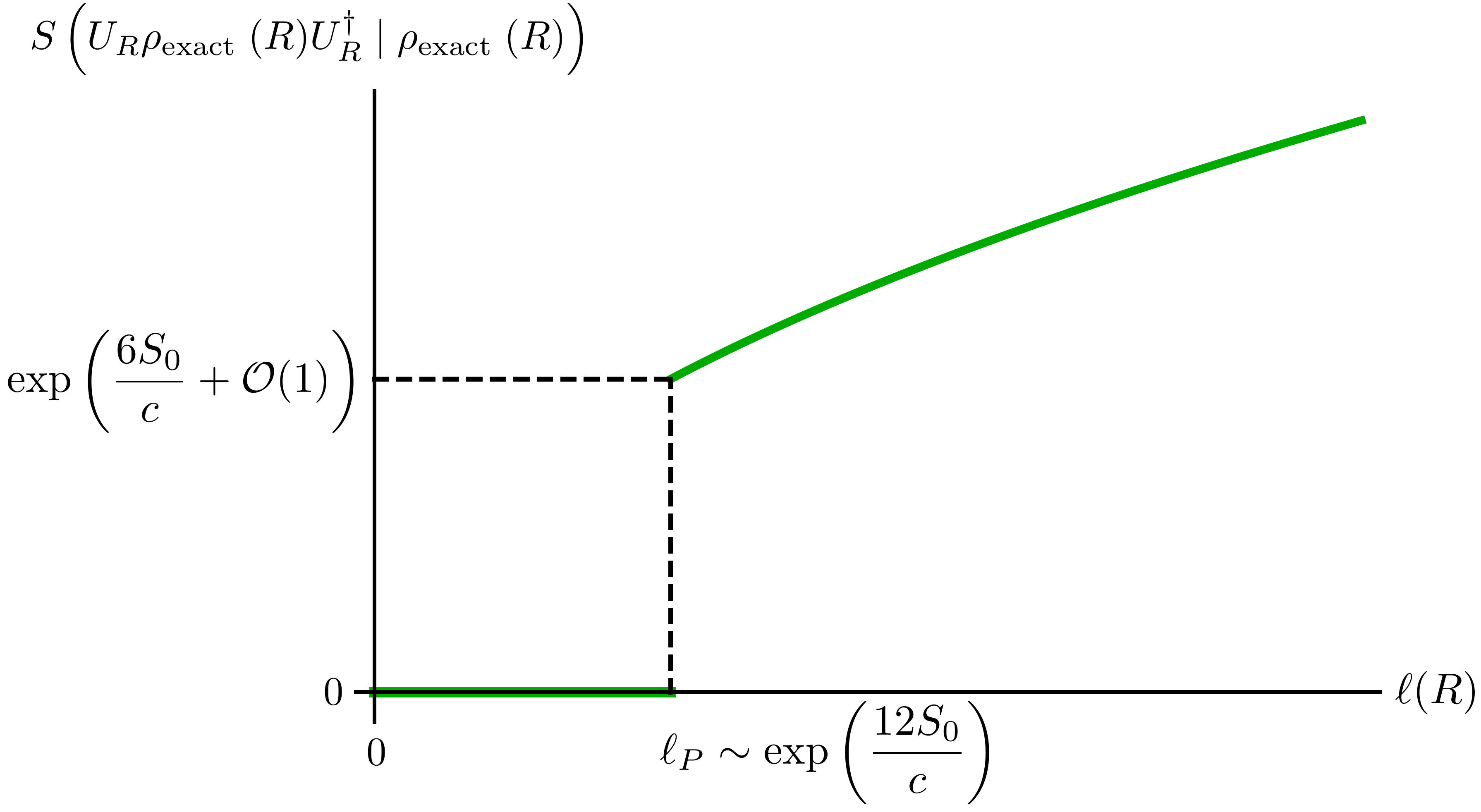}
\caption{The relative entropy in the free fermion model, as a function of the size of the region $R$ at fixed non-zero $\alpha$.
The horizontal axis is $\ell(R)$ the length of $R$ in units of $\phi_r/c$ ($\ell(R) \equiv c(b_2 - a_2)/\phi_r \approx c b_2/\phi_r$).
In the semi-classical approximation, there is a discontinuity in the relative entropy at the ``Page  length'' $\ell_P (R)\sim e^{12S_0/c}$ when an island appears in the gravity region, see (\ref{relresult1}).  It might seem that the relative entropy is large because it has $S_0$ in the exponential, but recall that in this model the ratio $S_0/c$ is kept finite and not very large \cite{Almheiri:2019yqk}, so the relative entropy is $\mathcal{O}(1)$.}
\label{fig:U1fermion}
\end{center}
\end{figure}

As discussed in \cite{Almheiri:2019yqk}, if we choose the bath region 
\begin{equation}\label{parameter1}
    a_2 \ll \frac{\phi_r}{c} \ll b_2, \quad \log \left( \frac{cb_2}{\phi_r}\right) > \frac{12 S_0}{c} + \mathcal{O}(1),
\end{equation}
then the island will be at
\begin{equation}\label{parameter2}
    a_1 \approx - b_2, \quad b_1 \approx - \frac{6\phi_r}{c}.
\end{equation}
Plugging (\ref{parameter1}) and (\ref{parameter2}) into (\ref{relresult}), we find
\begin{equation}\label{relresult1}
     S \left(  U_R \rho_{\textrm{exact}} (R)  U_R^\dagger \middle| \rho_{\textrm{exact}} (R) \right) \approx \frac{\pi}{4} \sin^2 \frac{\alpha}{2} \sqrt{\frac{cb_2}{3\phi_r}}.
\end{equation}
We observe that the relative entropy grows with the total central charge $c$, even though we are considering the symmetry that only involves one fermion. This confirms our previous expectation that the symmetry violation is contained in the correlation between the fermion field and other fields. We stress that $\rho_{\rm exact}(R)$ is the density matrix for all the fields, not just the one fermion field. Note that $\phi_r/c$ in (\ref{relresult1}) is roughly the time it takes for a particle to reach the island if it is released from the boundary. We will give a physical explanation of why this ``island time'' is relevant in \S\ref{increasing}. We sketch the result (\ref{relresult1}) in Figure \ref{fig:U1fermion}. There is an analogue of the ``Page time'' (perhaps more aptly called the Page length) 
$\ell_P$ in this problem that is the minimal size of $R$ to claim the island. Below the minimal size, the relative entropy is zero, while above that size, the relative entropy becomes finite and grows with the size. At finite $S_0$, we expect to have a smooth curve instead of a sharp transition, see \cite{Dong:2020iod,Marolf:2020vsi,Akers:2020pmf}.

\subsection{Increasing the relative entropy with charged particles \la{increasing} }

\begin{figure}[t!]
\centering
\includegraphics[width=15cm]{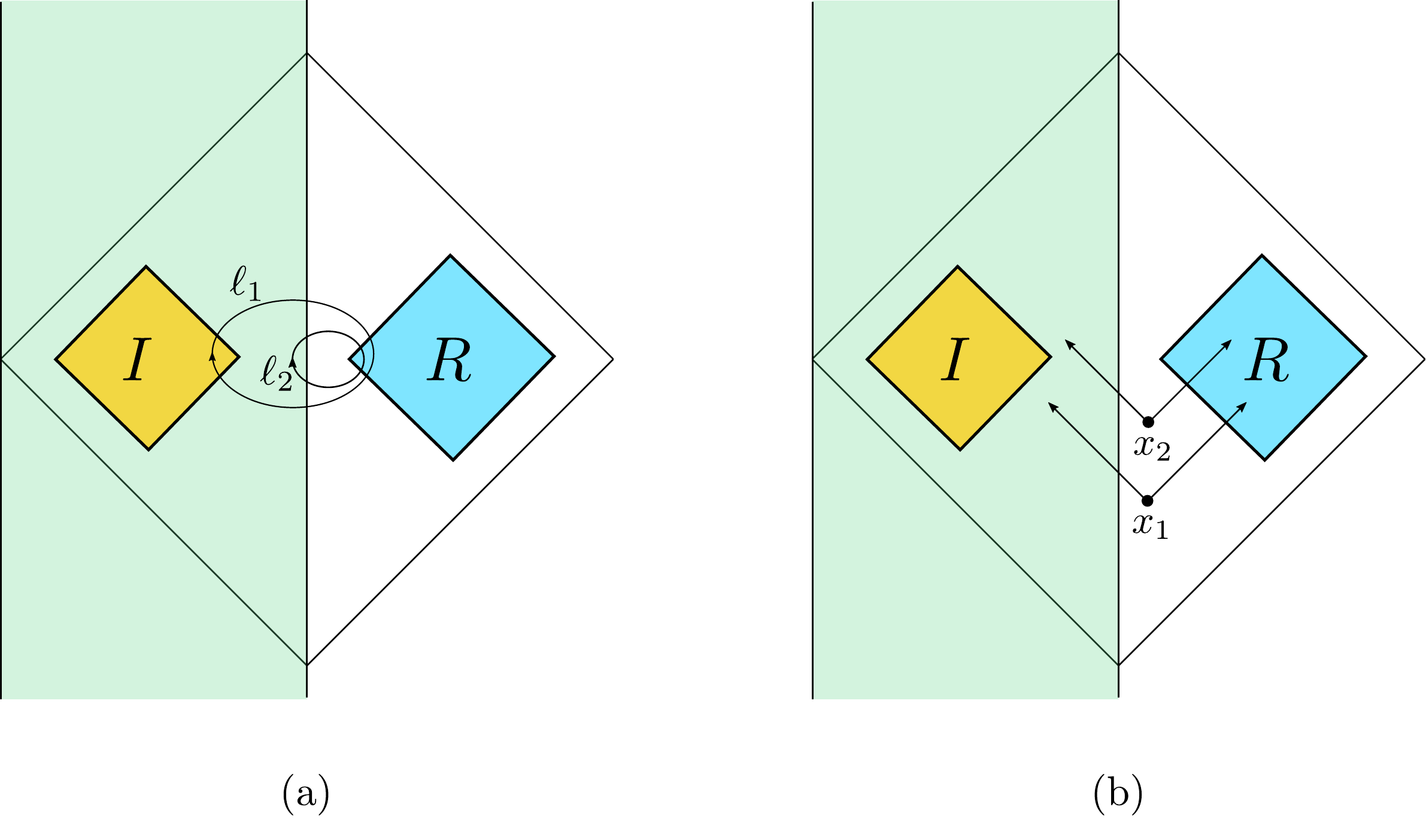}
\caption{(a) We consider two virtual processes $\ell_1$ and $\ell_2$ where charged particles run in loops. Only the larger loop $\ell_1$ contributes to the bulk relative entropy (as well as the entanglement between $I$ and $R$). (b) We imagine the decay of a neutral particle into 2 fermions at $x_1$ and $x_2$. Only the decay at $x_1$ contributes significantly to the bulk relative entropy. Note also that in an evaporating black hole, a morally similar picture would also predict that the relative entropy increases after the Page time because each Hawking mode will be entangled with a mode of opposite charge behind the horizon.
}
\label{bulk-rel-entropy}
\end{figure}

So far, we've only considered the ground state of the model in Figure \ref{fig:zerotemp}. It is natural to ask whether we can choose different bulk semi-classical states which will have a larger relative entropy when there is an island. To motivate the discussion, let us first look at a rough description of the semi-classical vacuum state. There are three regions, the island $I$, the radiation $R$, and the region in between --- call it $S$ for sea. The global state is charge neutral but locally there can be vacuum fluctuations of the charge density. We will focus on two possible forms of the fluctuation, and look at the density matrices they lead to. There are of course contributions from other types of terms and cross terms, but our goal here is just to draw some basic intuition. The first case corresponds to virtual loops that run between the sea and the radiation, as represented by $\ell_2$ in Figure \ref{bulk-rel-entropy} (a):
\eqn{ \ket{\rm vac}_\text{bulk} \sim \sum_q   \ket{0}_I \ket{-q}_S \ket{q}_R  , \la{psibulk1}}
where $Q_i \ket{q}_i = q_i \ket{q}_i, \, i= I,S,R$. The density matrix that it leads to is
\begin{equation}
    \rho_{\rm semi} (R\cup I) \sim \sum_q \left( \ket{0}_I  \bra{0}_I  \otimes \ket{q}_R \bra{q}_R  \right).
\end{equation}
It is invariant under $U_R$, and thus will not contribute to the relative entropy. The second case involves loops that run between the island and the radiation (loop $\ell_1$ in Figure \ref{bulk-rel-entropy} (a)):
\begin{equation}\la{psibulk2}
    \ket{\rm vac}_\text{bulk} \sim \sum_q   \ket{-q}_I \ket{0}_S \ket{q}_R ,
\end{equation}
and it leads to
\eqn{\rho_{\rm semi} (R\cup I) \sim  \sum_{q,\tilde{q}} \left( \ket{-q}_I \bra{-\tilde{q}}_I  \otimes \ket{q}_R  \bra{\tilde{q}}_R \right). \la{rhobulk}} 
We see that the semiclassical density matrix is not invariant under $U_R$, and will lead to a non-trivial dependence on $\alpha$ in the relative entropy via (\ref{relformula}). These loops are also the virtual processes that build up the entanglement between the island and the radiation, so we see that the relative entropy is closely related to entanglement, or mutual information. See Appendix \ref{app:subset} for some further discussion. 


The above toy model is supposed to depict the quantum fluctuations in the vacuum. However, it also suggests a way to increase the bulk relative entropy by making on shell particles with correlated charges. To neglect the fluctuations that already exist in the vacuum state, we could consider a matter theory which consists of a large $c$ CFT $+$ a massive ``probe'' theory with an $U(1)$ symmetry (for example, a single massive free fermion field). By ``probe'' field, we mean that when we can neglect the massive theory in determining the location of the island. When we compute the relative entropy with $U(1)$ symmetry generators acting on the massive theory using (\ref{DiffModular}), we expect it to be suppressed by $e^{-m \ell}$ where $\ell$ is the proper distance between the island and the bath region $R$, and is thus small for large mass. So if we create many pairs of particles, the leading contribution to the relative entropy will come from the real particles instead of vacuum fluctuations. In other words, we can interpret the charge fluctuations in (\ref{psibulk1}) and (\ref{psibulk2}) as coming from real particles.

Imagine that we start with a massive neutral particle at rest in the bath region. It then decays into two massive free fermions, with charge $+q$ and $-q$. These fermions propagate in opposite directions: one towards the black hole, and the other towards null infinity. However, we do not know which charge went where. 
It is interesting to consider the dependence of the relative entropy on the location $x$ of the fermion. If the particle which falls into the black hole is not in the island, then we have a state similar to (\ref{psibulk1}), and it will not increase the relative entropy significantly. However, if we drop in the particle and then wait some time $t_\text{island}$, the particle will appear in the island, see Figure \ref{bulk-rel-entropy}.
In cases where the island is near the horizon, for example in the evaporating black hole, or in the 2-sided setup at late times, we will need to wait about a scrambling time for a perturbation applied near the boundary of the gravity region.\footnote{In the 1-sided zero-temperature case, the scrambling time is infinite but the $t_\text{island} \sim \phi_r/c$. }
This will lead to a nonzero relative entropy, by considerations nearly identical to those in (\ref{psibulk2}) and (\ref{rhobulk}).



We would like to interpret the nonzero relative entropy in this case as saying that the charge is lost from the outside observer point of view. If the symmetry was exact, we should find a zero relative entropy at all time. The fact that we have a nonzero relative entropy after creating the particle pair simply means that the charge that is outside the region $R$ is not conserved. From our discussion, we see that this non-conservation happens at the time scale which is the scrambling time.
We will return to this point in the discussion.

\subsection{Evaporating black hole}


In the discussion of \S\ref{sec:general} and \ref{sec:example}, we mainly considered time independent cases where a black hole is in thermal equilibrium with the environment. However, our discussion also applies to general time dependent cases, such as evaporating black holes. Consider a theory that has a global symmetry in the semiclassical description, and we start from a pure state that has zero global charge and form a black hole. If the symmetry is not violated nonperturbatively, then if we look at the Hawking radiation emitted from the black hole, its density matrix should always be block diagonal in the charge basis, and thus is invariant under the symmetry transformation acting on the Hawking radiation. However, after the Page time, since the entanglement wedge of the Hawking radiation contains an island in the black hole interior, we can similarly compute relative entropies and see that the global symmetry is violated from a nonzero result. We also expect the relative entropy to grow as one collects more radiation from the black hole.

Of course, it is well known how to see that global symmetry is violated by looking at the final stage of evaporation as we reviewed in the introduction \cite{banks2011symmetries}. Here we are relating the violations of global symmetry to the appearance of an ``island.''
An advantage of the argument presented here is that it provides a quantitative way to see global symmetry violation just \emph{after the Page time}, and does not rely on the physics close to the final stage of the black hole evaporation. 

Notice that if we gather the entire radiation of a completely evaporated black hole, it will be in a pure state. The pure state will be a superposition of states with different charge. Hence we expect that the relative entropy after the black hole has completely evaporated to formally diverge. On the other hand, the R\'{e}nyi versions of the relative entropy that we study in \S\ref{sec:replica} will still be well defined and give sensible answers. A related point is discussed in Appendix \ref{app:westcoast}, where we explore global symmetry violations in a different model of JT gravity.

Instead of starting with a state that is annihlated by the charge, one could also imagine forming the black hole from a general initial state $\rho_0$ that is not invariant under the global symmetry $U$ that acts on the entire system.\footnote{Here we are assuming that before the black hole forms, all the matter are far apart and the gravity effect is weak enough such that the operator $U$ is well defined.} In this case, one could still see that global symmetry is violated as follows. If the symmetry were exact, then when we look at any subsystem $R$ at later time $t$, we should always find
\eqn{ \sr{ U_R \rho_R(t) U_R^\dagger}{\rho_R(t) } & \le \sr{ U e^{i H t} \rho_0 e^{-i H t} U^\dagger}{e^{i H t} \rho_0 e^{-i H t}}\\
& =S(U\rho_0 U^\dagger |\rho_0), \la{inequality} }
where on the first line we are using the monotonicity of relative entropy under restricting to a subsystem $R$, and we used the assumption that $U$ is an exact symmetry to get to the second line. However, if we use the gravitational formula to compute $\sr{ U_R \rho_R(t) U_R^\dagger}{\rho_R(t) }$, we expect to get an increasing function with $t$ after the Page time,\footnote{We expect the curve to be qualitatively similar to Figure \ref{fig:U1fermion}, if we replace the size of the interval by time $t$. The reason is that there will be more and more correlated charge fluctuations between $I$ and the Hawking radiation $R$ (this is similar in spirit to Figure \ref{bulk-rel-entropy}). It would be interesting to understand the calculation for evaporating black holes in more detail. } so it cannot be bounded by a constant $S(U\rho_0 U^\dagger |\rho_0)$ as in (\ref{inequality}), which necessarily means that the global symmetry is violated.

\section{Global symmetry violation and replica wormholes}\label{sec:replica}

\subsection{Charge flowing through the replica wormhole}\label{sec:wormhole}

\begin{figure}[t!]
\begin{center}
\includegraphics[width=7.5cm]{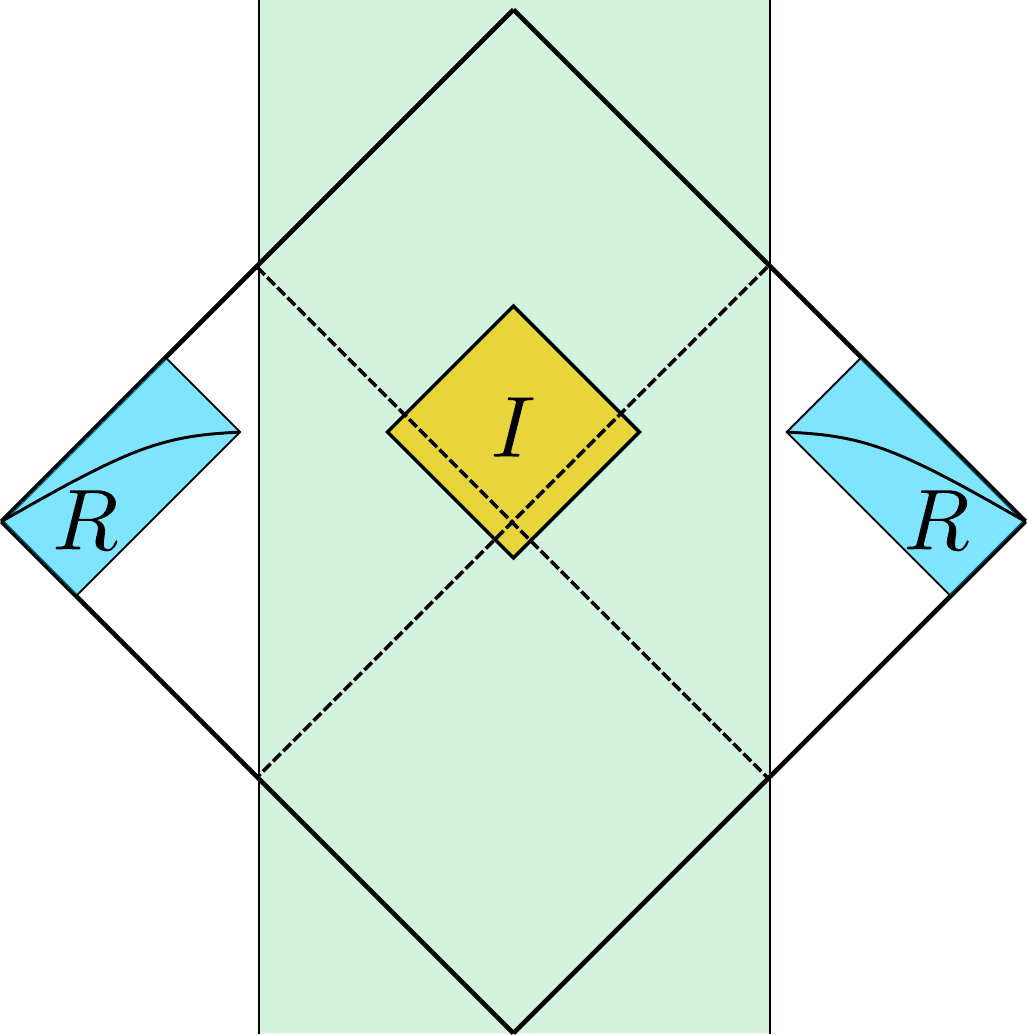}
\caption{A two-sided black hole in JT gravity coupled to two bath regions \cite{Almheiri:2019qdq}. When we compute the entropy of a large region $R$ in the bath, depending on the parameters, we can have an island $I$ in the gravity region.
At sufficiently late times $t>0$ there will always be an island.
}
\label{fig:twosided}
\end{center}
\end{figure}

As we've seen in \S\ref{sec:relentropy}, when there is an island in the entanglement wedge of $R$, the global symmetry violation is reflected in the relative entropy. In the computation of the von Neumann entropy using the replica trick, the island arises from the replica wormholes \cite{Almheiri:2019qdq,Penington:2019kki}.
So it is natural to consider the replica version of the relative entropy, and ask what the role of replica wormholes are in these quantities.
Similar to the von Neumann entropy, the relative entropy can be computed via a replica trick \cite{Lashkari:2014yva} by
\begin{equation}
    S(\rho|\sigma) = \lim_{n\rightarrow 1} \frac{1}{1-n} \tr \left[ \rho (\sigma^{n-1} - \rho^{n-1}) \right]. \la{renyilike}
\end{equation}

To be concrete, we will take the two-sided black hole setup in \cite{Almheiri:2019yqk} as an example, whose replica wormhole solution has been constructed explicitly in certain limits in \cite{Almheiri:2019qdq}. The goal of this section is not to study the replica wormhole geometries in detail, but to understand what features of the replica wormholes are related to charge violation.
We will be considering the thermofield double state $\ket{\rm{TFD}}$ of the system, and study the density matrix of a region $R$ which contains part of both the left bath and the right bath (see Figure \ref{fig:twosided}). Depending on the parameters, we can have an island $I$ that is slightly outside the horizon already at $t=0$; our arguments in \S\ref{sec:relentropy} would then apply.

\begin{figure}[t!]
\begin{center}
\includegraphics[width=15cm]{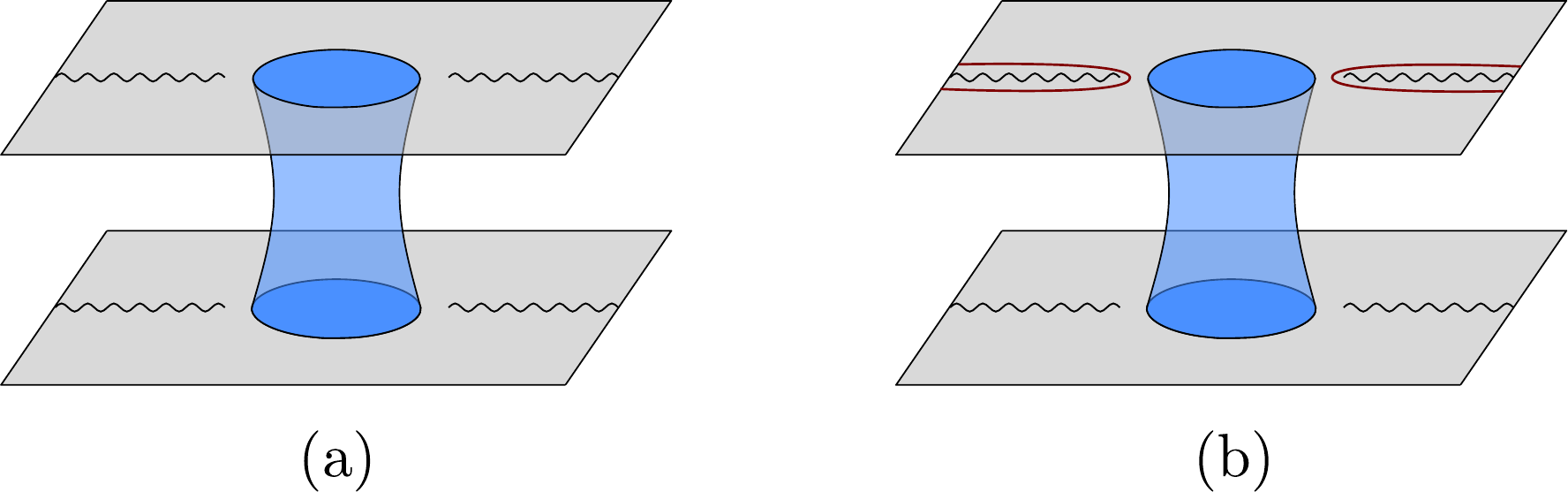}
\caption{(a) The replica wormhole geometry $\mathcal{M}_2$ for computing $Z_2 (0)$. (b) For $Z_2 (\alpha)$, we have the symmetry transformation operators inserted around the cuts in one of the replicas. }
\label{rep1}
\end{center}
\end{figure}
Here we would like to consider the $n=2$ replica:
\begin{equation}\label{Z2alpha}
    \frac{Z_2 (\alpha)}{Z_1^2} =  \textrm{tr}\left[U_R \rho_{\rm exact} (R) U_R^\dagger \rho_{\rm exact} (R)  \right], \quad U_R = e^{i\alpha Q_R},
\end{equation}
where we take the symmetry to be an $U(1)$ global symmetry just for simplicity. We also divided by $Z_1^2$ where $Z_1$ is given by the path integral on a single replica, to get the expression for a normalized density matrix. If the symmetry were exact, then we should have $Z_2 (\alpha) = Z_2 (0)$, for the reasons discussed in \S\ref{sec:general}. As explained in \cite{Almheiri:2019qdq}, $Z_2 (0)$ can be computed via a gravitational path integral with two replicas, and in the case with an island, the path integral is dominated by a replica wormhole geometry (see Figure \ref{rep1} (a)), which we denote by $\mathcal{M}_2$. The only difference in the computation of $Z_2(\alpha)$ is that we have the extra insertions of $U_R$ and $U_R^\dagger$, which are represented by the red lines in figure \ref{rep1} (b).  

\begin{figure}[h]
\begin{center}
\includegraphics[width=15cm]{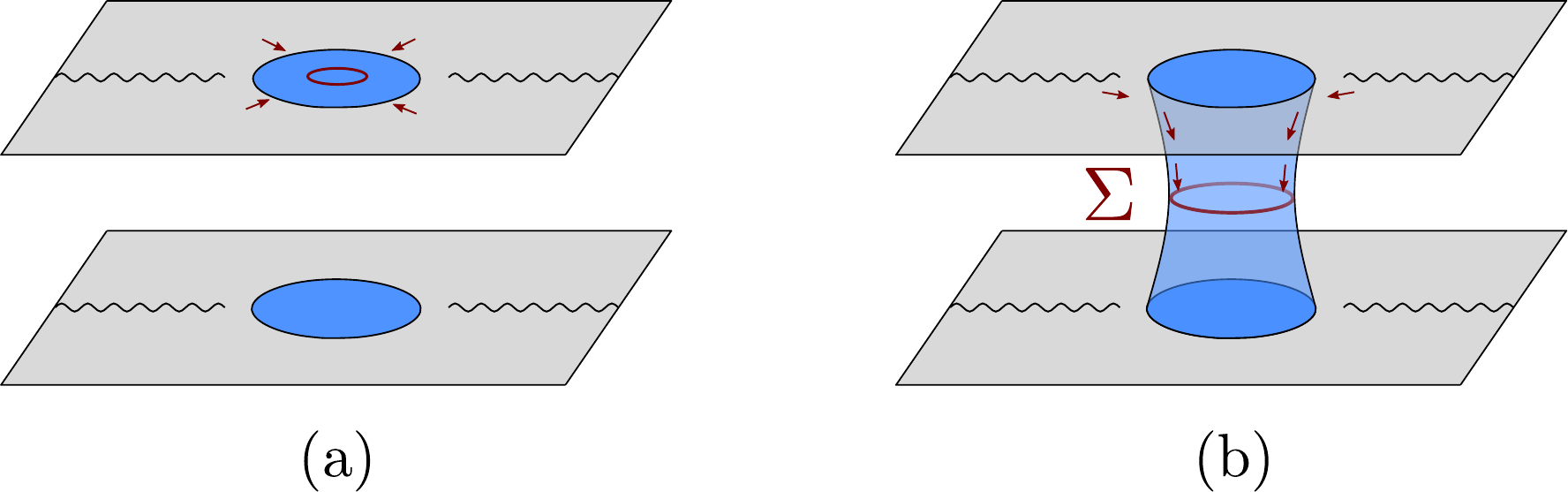}
\caption{(a) Without the replica wormhole, the topological operator can be shrunk to a point. (b) With the replica wormhole, we can deform the topological operator to surround the throat of the wormhole. Therefore the dependence on the parameter $\alpha$ diagnoses the charge flowing through the wormhole. }
\label{rep1a}
\end{center}
\end{figure}

Importantly, since we are computing (\ref{Z2alpha}) in the semiclassical description that has the global symmetry, the red lines in Figure \ref{rep1} (b)  represent topological operators.\footnote{The topological operator is well-defined for other symmetries as well, including non-Abelian cases and discrete symmetries.} This means we can deform the curves arbitrarily as long as they do not cross any charged operators (in which case they would pick up a phase).
Since we are considering the thermofield double state, which does not contain any insertions, we can deform the topological operator freely. If there were no replica wormholes (see Figure \ref{rep1a} (a)), we could then deform and shrink the topological operator to a point, which tells us
\begin{equation}
    \frac{Z_2 (\alpha)}{Z_2 (0)} = 1, \quad (\textrm{no replica wormhole}).
\end{equation}
On the contrary, with the replica wormhole, the topological operator cannot be shrunk to a point, but it can be deformed to the throat $\Sigma$ of the wormhole, see Figure \ref{rep1a} (b). So we have
\begin{equation}
    \frac{Z_2 (\alpha)}{Z_2 (0)} = \langle  e^{i\alpha Q_{\rm wormhole}}  \rangle_{\mathcal{M}_2}, \quad (\textrm{with replica wormhole}),
\end{equation}
where $Q_{\rm wormhole} = \int_{\Sigma} *J$ and $J^\mu$ is the Noether current.
This is the finite replica analogue of equation (\ref{relEntropy2}). 
In general, we expect that $\langle  e^{i\alpha Q_{\rm wormhole}}  \rangle_{\mathcal{M}_2} < 1$, simply because the value of $Q_{\rm wormhole}$ can fluctuate in the path integral. 
Thus we would have
\begin{equation}
    Z_2 (\alpha) < Z_2 (0), 
\end{equation}
which means that the density matrix $\rho_{\rm exact} (R)$ is not invariant under the symmetry transformation, as we concluded by studying the relative entropy.

\begin{figure}[h]
\begin{center}
\includegraphics[width=15cm]{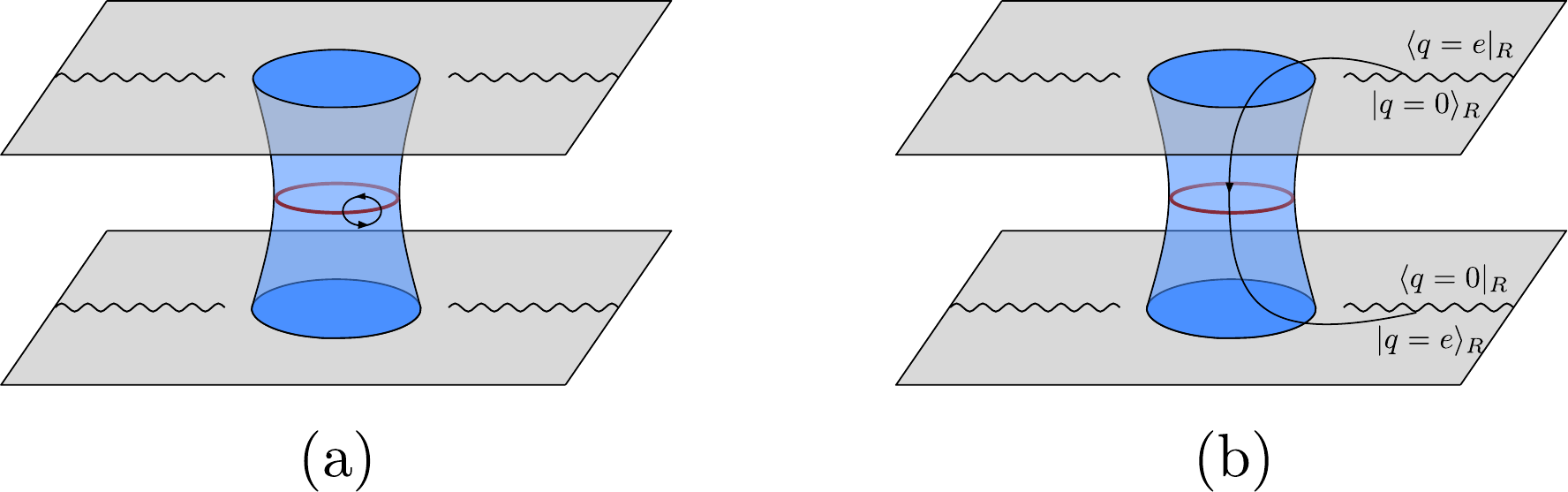}
\caption{ Two virtual processes that contribute to $Z_2(\alpha)$. (a) The smaller, contractible loop does not give a contribution that depends on $\alpha$, since on the red circle there is both a particle and an anti-particle. (b) The larger loop is non-contractible, contributes to the off-diagonal part $\textrm{tr}\left[\rho_{e,0}\rho_{e,0}\right]$ (see \S\ref{sec:offdiagonal}), and therefore to the $\alpha$ dependence. This should be compared with Figure \ref{bulk-rel-entropy}.}
\label{virtual}
\end{center}
\end{figure}

We can use the intuition from a weakly-coupled theory to understand this effect slightly better. If we consider a small (meaning contractible) loop of virtual particles near the throat of the geometry, the net charge flow will be zero, since on a spatial slice $\Sigma$, we will have one particle and one anti-particle, see Figure \ref{virtual} (a). However, the virtual loop could be non-contractible due to the cuts in the bath region, see Figure \ref{virtual} (b).

\subsection{Off-diagonal contributions to the density matrix}\la{sec:offdiagonal}
The fact that $U_R \rho_{\rm exact} (R) U_{R}^\dagger$ and $\rho_{\rm exact}(R)$ are different indicates that $\rho_{\rm exact}(R)$ must have some off-diagonal elements in the charge basis. 
We can try to ``open up" the quantity in (\ref{Z2alpha}) and look at the contribution to the off-diagonal elements more explicitly:
\begin{equation}\label{offdiag}
\operatorname{tr}\left[ e^{i \alpha Q_{R}} \rho_{\rm exact} (R) e^{-i \alpha Q_{R}} \rho_{\rm exact} (R) \right] =\sum_{q}\operatorname{tr}\left[ \rho_{q,q}^2 \right]  +  e^{i\alpha(q-q')}\sum_{q\neq q'}\operatorname{tr}\left[ \rho_{q,q'} \rho_{q' ,q}\right],
\end{equation}
where
\begin{equation}
    \rho_{q,q'} \equiv P_q  \rho_{\rm exact} (R) P_{q'}
\end{equation}
and 
$P_q \equiv \sum_{i} \ket{q,i} \bra{q,i}$ is the projector into the subspace with $Q_R = q$, with $i$ running over the states in the same-charge subspace. Note that the projection only acts on the fields that transform under the symmetry, which we take to be a small subset of the fields. Since we are tracing out all the other fields, we expect a similar replica wormhole solution for each term in (\ref{offdiag}) independently. We see from (\ref{offdiag}) that all the dependence on $\alpha$ comes from the off-diagonal terms in the density matrix. 
In this section, we give a more direct explanation of the appearance of off-diagonal terms, and how they arise from Euclidean wormholes in gravity.

One way to understand the off-diagonal term $\operatorname{tr}\left[ \rho_{q,q'} \rho_{q' ,q}\right]$ is as follows. We can rewrite the quantity as
\begin{equation}
  \operatorname{tr}\left[ \rho_{q,q'} \rho_{q' ,q}\right]   = \sum_
  {i,j} |\braket{\textrm{TFD} | q, i} \braket{ q', j|\textrm{TFD}}  |^2,
\end{equation}
and in the path integral, it corresponds to putting boundary conditions for the fields along the branch cuts such that they have definite charges. (This should not be taken too literally, as such a state might be a very high energy.) In the semiclassical approximation, we can think of the boundary conditions as places where the worldline of a charged particle can be created or destroyed. Let's take $\operatorname{tr}\left[ \rho_{e,0} \rho_{0 ,e}\right]$ as a simple example, where $e$ is the unit charge of a charged particle. Due to the boundary condition, we will have a worldline starting from the bra part of the cut in the first replica, and the only place where it can land on is the ket part of the cut in the second replica. The only possibility for the contribution to be nonzero is that the charge can go through the replica wormhole, as shown in Figure \ref{virtual} (b). 

Despite that we have $\operatorname{tr}\left[ \rho_{q,q'} \rho_{q' ,q}\right] \neq 0$, if we compute $\operatorname{tr}\left[ \rho_{q,q'}\right]$ with $q\neq q'$ using gravity, we will find a zero answer, up to nonperturbative corrections. This is because we only have one replica, and there is nowhere the charged particle can go to. 
This effect can be viewed as a special case of a more general phenomena involving Euclidean wormholes. Consider a state of the effective quantum field theory in the bath obtained by inserting some charged operators in the thermofield double state $\ket{\psi_q} = O_{q_1} O_{q_2} \cdots O_{q_n} \ket{\textrm{TFD}}$. Here $O_{q} = O_q(\tau,x)$ is some local operator with charge $q$ evaluated at a general Euclidean or Lorentzian time.
Ignoring the effects of higher topologies, this state will be orthogonal to the state $\ket{\psi_{\tilde{q}}} = O_{\tilde{q}_1} O_{\tilde{q}_2} \cdots O_{\tilde{q}_m} \ket{\textrm{TFD}}$ unless the total charges are equal $q = \sum_i q_i =  \tilde{q} = \sum_i \tilde{q}_i$, namely $\braket{\psi_q | \psi_{\tilde{q}}} = 0$ up to exponentially small errors. 
However, wormhole geometries will lead to a small squared-overlaps when $q\neq \tilde{q}$ \cite{Stanford:2020wkf}:
\begin{equation}\la{charge-worm}
    | \braket{\psi_q|\psi_{\tilde{q}}} |^2  = \delta_{q-\tilde{q}}  \lp \textrm{disconnected} \rp + e^{-S_0}  (\textrm{connected})
\end{equation}
The connected contribution comes from an Euclidean wormhole, where the charged particles go through the wormhole. For example, the square of a 1-pt function $\ev{O_q}$ with charge $q$ is given by
\eqn{|\ev{\textrm{TFD}\middle| O_q \middle| \textrm{TFD}}|^2 \quad  = \quad \includegraphics[scale = 0.7, valign=c]{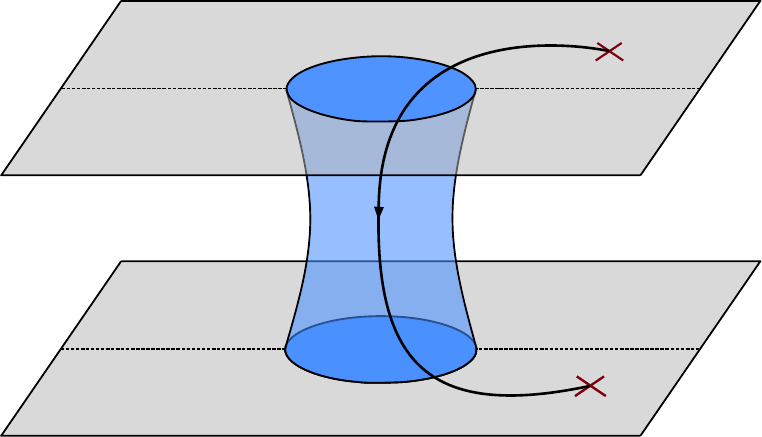}.\la{OqWormhole} }
Note however that (\ref{OqWormhole}) might require an ensemble average interpretation, since the left hand side is the square of $\bra{\textrm{TFD}}O_q \ket{\textrm{TFD}}$, while the right hand side does not factorize manifestly. In a single unitary system, the proper interpretation of (\ref{OqWormhole}) might be that the wormhole computes the average of $|\bra{\textrm{TFD}}O_q \ket{\textrm{TFD}}|^2$ among a suitable family of operators with charge $q$. We refer the readers to \cite{Maldacena:2004rf,Cotler:2016fpe,Saad:2018bqo,Saad:2019lba,Saad:2019pqd,Stanford:2020wkf,Pollack:2020gfa,Marolf:2020xie,Giddings:2020yes,McNamara:2020uza,Bousso:2020kmy} for more discussions on these issues about Euclidean wormholes. We should stress that the quantity $\textrm{tr}\left[ \rho_{q,q'} \rho_{q',q}\right]$ does not factorize in the first place, since each charge subspace still contains a huge number of states that we sum over. So the quantity we compute does not suffer from factorization problem immediately. Due to the summation over a large number of intermediate states, we have a replica wormhole geometry which is a saddle point solution that dominates the path integral, while in general the wormhole in (\ref{OqWormhole}) will be an off-shell configuration (see however an on-shell construction in \cite{Stanford:2020wkf}).

\begin{figure}[t!]
\begin{center}
\includegraphics[width=6cm]{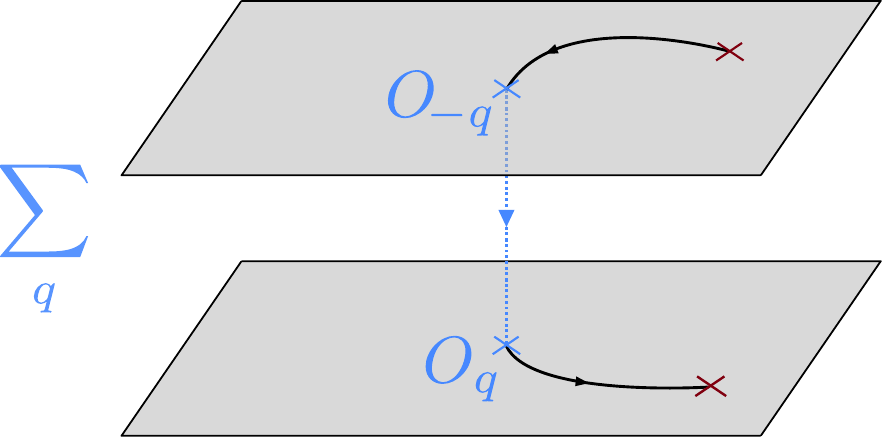}
\caption{We can reproduce the effects of the wormhole by an effective picture where we sum over correlated operators. This is also what happens from the boundary point of view, where the gravity region is replaced by a boundary condition which completely breaks the global symmetry. Here we are imagining that the operator insertions in the bath are far away from the gravity region. In some theories with a disorder average, the dashed blue line may represent a contraction of couplings.}
\label{dual}
\end{center}
\end{figure}

If we imagine that the wormhole region is far away from the operator insertions, we can summarize the effects of the wormhole using an effective picture. If we think of the gravity region as a complicated boundary condition for the bath fields, this boundary condition will break explicitly the global symmetry. If the gravity theory has a holographic description as we've assumed in \S\ref{sec:relentropy}, then this boundary condition would be realized by the coupling between the quantum mechanical system and the bath CFT. More generally, we expect the same picture to hold even when the gravity theory does not have a holographic description. Far away from this boundary condition, the circle is hard to distinguish from a point, so we can replace the boundary condition by a sum of operator insertions, see Figure \ref{dual}. We can imagine that there is a small coefficient for some operator $c_q O_q$ where $c_q$ has mean zero in some appropriate averaged sense. Then ignoring these operators is equivalent to only considering disconnected geometries. When we include connected geometries, we are taking into account the non-zero variance of these coefficients $|c_q|^2 \sim e^{-S_0} $. Of course, this is just the old story about ``$\alpha$-parameters" and Euclidean wormholes \cite{Coleman:1988cy,Giddings:1988cx,Giddings:1988wv,Marolf:2020xie}, what we did in this section is making the relation to replica wormholes more explicit.

From this perspective, it is clear that a Euclidean wormhole with charge $\pm q$ propagating through the throat is giving rise to a state with charge $\pm q$ different than the ``naive'' charge of the state (e.g. the charge of the state we would have assigned in the effective description of the quantum fields on the trivial topology).

\begin{figure}[h]
\begin{center}
\includegraphics[width=15cm]{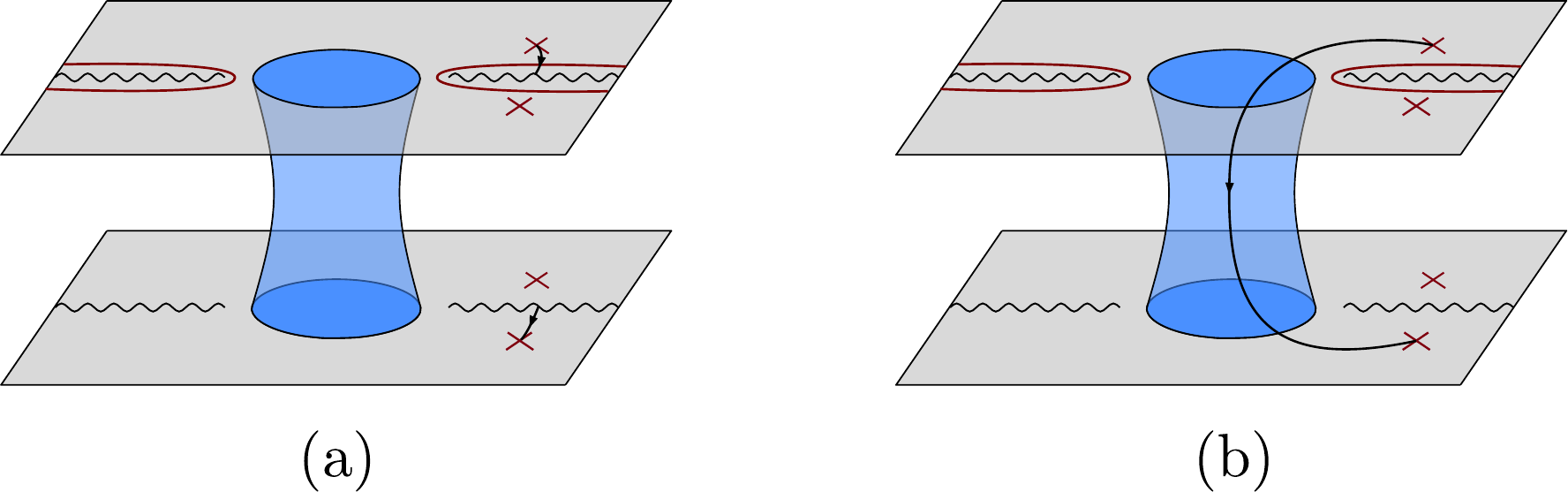}
\caption{We consider a global state which is not the vacuum but contains some operator insertions in the bath region. (a) The diagonal contribution to the density matrix comes from particles propagating through the cuts. (b) Off-diagonal contributions come from world-lines with non-zero net charge propagating through the wormhole. }
\label{excited}
\end{center}
\end{figure}


Now let us return to the density matrix of the fields in a subregion $R$.
We will slightly generalize the previous discussion by considering a state $\ket{\psi_q}$ which is obtained by inserting some charged operators in the Euclidean evolution. Let us denote elements of a charge eigenbasis by a composite index $r$ such that $\ket{r} = \ket{q_r, i}$, where $i$ runs over states with the same charge $q_r$.
Then, 
\eqn{ \rho_{\rm exact} (R)  &= \tr_{\bar{R}} \ket{\psi_{q}}\bra{\psi_{q}},\\
Z_2(\alpha)/Z_1^2 &= \sum_{r,s} \bra{r} e^{-i \alpha Q} \rho_{\rm exact} (R)  e^{i \alpha Q}\ket{s} \bra{s} \rho_{\rm exact} (R)  \ket{r}  \\
& = \sum_{r,s} e^{i\alpha (q_s - q_r)} \bra{\psi_{q}} O_{s,r}^R \ket{\psi_{q}}\bra{\psi_{q}} O_{r,s}^R \ket{\psi_{q}}   \\
 & = \sum_{r,s} e^{i\alpha (q_s - q_r)} |\braket{\psi_q |\psi_{\tilde{q}}}|^2\\
 \ket{\psi_{\tilde{q}}} &=  O_{r,s}^R \ket{\psi_{q}}, \quad  \tilde{q} = q+q_r -q_s. }
where $O^R_{r,s} \equiv \ket{r}\bra{s}$ is an operator with charge $q_r - q_s$.
Neglecting gravity, we would have expected to only receive a contribution when $q_r = q_s$. However, from \nref{charge-worm} we see that each off-diagonal contribution to the wormhole corresponds to diagrams where particles carrying total charge $q_r- q_s$ propagate through the wormhole, see Figure \ref{excited}.

As a side remark, let us comment on the relation between scattering in the black hole background.
We have seen that the same effect which gives the non-zero relative entropy is also responsible for non-zero values of correlation functions which would naively vanish from symmetry considerations. These correlation functions in the bath are closely related to scattering in the black hole background. For example, a 4-pt function in the bath region in the appropriate out-of-time-order configuration could be interpreted as the traversable wormhole signal \cite{Gao:2016bin} that violates global symmetry. 

\subsection{Comments on gauge symmetries}\label{sec:gauge}

While exact global symmetries are not allowed in quantum gravity, gauge symmetries of course can exist.
Here we explain the difference between the two from the replica wormhole point of view. To be concrete, we will consider the same set up as in (\ref{sec:wormhole}), where the CFT in the flat space region has an $U(1)$ global symmetry, while the difference is that the symmetry is gauged in the gravity region.
By the same consideration as in (\ref{sec:wormhole}), we would still have
\begin{equation}
    \frac{Z_2 (\alpha)}{Z_2 (0)} = \langle  e^{i\alpha Q_{\rm wormhole}}  \rangle_{\mathcal{M}_2}, \quad (\textrm{with replica wormhole}).
\end{equation}
However, in the path integral, there will be an integral over the spatial zero mode of the gauge field
$\int dA_0 \exp \lp {i A_0 \int_\Sigma j^0 } \rp = \delta(Q_{\rm wormhole}) $
which sets the charge propagating through the wormhole to zero. Thus we have
\begin{equation}
    Z_2 (\alpha) = Z_2 (0)
\end{equation}
for arbitrary $\alpha$ as enforced by the gauge constraint. Similarly we have $Z_n (\alpha) = Z_n (0)$ for higher replicas as well. 
It then follows that as $n\to 1$ the relative entropy must vanish. This shows that the symmetry is exact in the boundary description as expected.\footnote{It would be interesting to understand how to see the vanishing of the relative entropy directly from the relative entropy formula (\ref{relEntropy}), without going through the replica argument.}

We can rephrase this argument slightly differently as follows. Consider $(d+1)$-dimensional electromagnetism on a compact space with no boundary $\Sigma \times$ time. In such a theory, it is clear that the overall charge on $\Sigma$ must vanish. A net charge would source electric field lines, which have nowhere to go. Integrating Gauss's law
$\nabla \cdot E = \rho $
on a spatial manifold $\Sigma$ with no boundary, we get the constraint that $Q= 0$.
If we think of the direction along the Euclidean wormhole as time, this makes it clear that no net charge can propagate through the wormhole. This is true even when the gauge coupling goes to zero. Note that if we view gravity as a gauge theory, the analogous equation says that the ADM Hamiltonian of a closed universe vanishes. If we perform a Kaluza-Klein reduction of some of the compact dimensions along the wormhole, the momenta in the compact dimensions will become gauge charges. So we will get a constraint that the total momentum in each compact dimension vanishes.

\section{Discussion and conclusion}\la{sec:conclusion}



It has been suggested for a long time that Euclidean wormholes lead to symmetry violation since charged particles may propagate through the wormhole. 
Normally, this is an exponentially small effect. The main point of this paper is showing that when we compute certain relative entropies, the tiny effects add up and give an $\mathcal{O}(1)$ result. 
This is morally similar to what happens in the Page curve after the Page time, where the tiny correlations in the Hawking radiation bring down the entropy. The novelty here is that instead of computing the entropy, we are computing a relative entropy which involves symmetry generators acting in the bath region, so these quantities tell us about the symmetry violation. Importantly, this effect can be computed with just knowledge of the semi-classical description. Our results are based on the recent paradigm of islands and replica wormholes. The same logic which demands that replica wormholes be included in the computation of the Page curve also demands that they should be included in computations of the relative entropy. Of course, there are unresolved puzzles about the inclusion of Euclidean wormholes in gravitational path integrals \cite{Maldacena:2004rf,Cotler:2016fpe,Saad:2018bqo,Saad:2019lba,Saad:2019pqd,Stanford:2020wkf,Pollack:2020gfa,Marolf:2020xie,McNamara:2020uza,Bousso:2020kmy}. In some cases, it is known that Euclidean wormholes are computing a disorder average over theories. In these cases, it is possible that the ensemble may have a symmetry group $G$ that each individual theory does not. (This means that the probability distribution over theories is invariant under the action of $G$ on theory-space). It is tempting to say that when there is an ensemble interpretation, if we have a matter theory with global symmetry group $G$ coupled to gravity, the group $G$ should be re-interpreted as the symmetry of the ensemble or the third-quantized theory. 
See Appendix \ref{app:C3} for some extra comments.

The relative entropy $S(U_R \rho_\text{exact} (R) U_R^\dagger | \rho_\text{exact} (R))$ that we considered involves only the density matrix in the bath region. Although it is not a direct observable, it might be able to provide contraints or bounds on physical observables, which requires further investigation.  
If one can create (or simulate holographically) black holes in the lab, then the relative entropy and its R\'{e}nyi versions are in principle measurable for observers who have multiple copies of the system.
In principle, measuring the say $n=2$ version of the relative entropy would not be hard (in the sense of computational complexity) if one had access to $n=2$ copies of the system. We do not need to measure each component of the density matrix independently.


We stress that it is surprising that a quantity that characterizes the nonperturbative violation of global symmetry can be computed in the semiclassical description. This suggests that in some sense, the way gravity violates global symmetry is far from generic. By this we mean the following: if we take some quantum field theory and put a random boundary condition that contains tiny terms that breaks any global symmetries, we would not expect the exact relative entropy to satisfy any formula given by the effective theory where we have erased the tiny terms. Of course, the surprise we are referring to is similar to the surprise that semiclassical gravity knows about the Page curve.

We've focused on idealized situations where there is no global symmetry violation within the semiclassical description. However, the argument we had can be easily generalized to situations where the symmetry has already been violated semiclassically. (For example, there could be Planck-suppressed operators in the Lagrangian which break explicitly any global symmetries.)
In such theories, the relative entropy of the semiclassical density matrix will already be non-zero, i.e. $ S \left(U_R \rho_{\textrm{semi}} (R)  U_R^\dagger \middle| \rho_{\textrm{semi}} (R) \right)  >  0$. However, if we consider the relative entropy of the exact density matrix, as long as the island still exists, we will have
\begin{equation}
\begin{aligned}
       S \left(  U_R \rho_{\textrm{exact}} (R)  U_R^\dagger \middle| \rho_{\textrm{exact}} (R) \right) & =    S \left( U_R \rho_{\rm semi} (R \cup I)  U_R^\dagger \middle|  \rho_{\rm semi} (R \cup I)\right) \\
      & >   S \left(U_R \rho_{\textrm{semi}} (R)  U_R^\dagger \middle| \rho_{\textrm{semi}} (R) \right),
\end{aligned}
\end{equation}
where we used the monotonicity of relative entropy from the first line to the second line. The difference between the exact answer and the semiclassical answer will still be order one due to the inclusion of the island. Thus for an evaporating black hole, we will still find a sudden increase of the relative entropy at the Page time, which is a clear signature of the global symmetry violation from nonperturbative effects.

%
We have emphasized that the symmetry violating effects we are considering are related to Euclidean wormholes. On the other hand, it has long been known that the Hawking process violates symmetries; these previous arguments only involved the naive disconnected geometries.
In order to emphasize what our new arguments buy, let us return to the setup discussed in \S \ref{increasing}. The matter theory in the bulk is a large $c$ CFT $+$ a probe massive field with a $U(1)$ symmetry and particles of mass $m$. Let us consider a large black hole which has a Hawking temperature $T \ll m$ that is older than the Page time. Imagine that it formed from neutral matter. Now consider the same experiment in \S\ref{increasing}. If a few charged particles fall into the black hole, we might wonder whether the experimentalist which remains outside the hole can see charge violation. Naively to check the argument of \cite{banks2011symmetries}, she would have to wait an extremely long time for the black hole to evaporate. In particular, she would have to wait for the black hole to become small enough that $T \gg m$, at which point it could start to radiate the $U(1)$ charge. 

However, our proposal is that even after a scrambling time, the clever experimentalist can in fact report a violation of charge conservation by measuring off-diagonal components of the density matrix, or if she can control multiple copies of the system, a suitable R\'{e}nyi relative entropy.

Note that our result is similar in spirit to the result of Hayden and Preskill \cite{Hayden:2007cs}. Naively, one would have guessed that to recover the information from a diary thrown into the hole, one would have to wait until almost the end of the evaporation process. 
But \cite{Hayden:2007cs} showed that one only needs to wait for a scrambling time if the black hole is older than the Page time. Here we are in some sense refining their result and showing that the information in the Hawking radiation recovered after the scrambling time is sufficient to demonstrate that charge conservation is violated.\footnote{Despite the analogy, we should not confuse the ability to recover the charge of the particle using entanglement reconstruction with the effect that we are talking about here. The violation of charge conservation is more directly related to the fact that naively orthogonal black hole states are not orthogonal in the exact description, this is particularly clear in the model of Appendix \ref{app:westcoast}.}

The connection between the appearance of islands and violation of global symmetries implies that we can see clear signatures of symmetry violation even in situations where charge conservation considerations like those in \cite{banks2011symmetries} do not immediately apply. Besides the Hayden-Preskill-like experiment discussed above, an example is the extremal (zero temperature) black hole. Another case is the finite temperature black hole in equilibrium with a bath that never evaporates. We expect our results to generalize to the cosmological setups of  \cite{Chen:2020tes,Hartman:2020swn,Balasubramanian:2020hfs} where the old arguments would not establish a clear signature of symmetry violation. Of course, the old arguments only involve measuring simple observables like the total charge of the Hawking radiation. In our work, we are discussing the relative entropy which is a much more complicated quantity. 

It would be interesting to see if our arguments can be generalized to address questions about gauge charges like the weak gravity conjecture, completeness conjecture, etc.  \cite{Polchinski:2003bq, ArkaniHamed:2006dz}.




\paragraph{Acknowledgement}
We would like to thank Juan Maldacena for helpful discussions during the project. We thank Juan Maldacena and Daniel Harlow for useful comments on a draft of the manuscript. We also want to thank Ahmed Almheiri, Yingfei Gu, Ho Tat Lam, Edgar Shaghoulian, Leonard Susskind, Edward Witten, Pengfei Zhang for interesting discussions.

\appendix



\section{Nonperturbative correlation between subsets of fields in the Hawking radiation}\label{app:subset}

As we mentioned in \S\ref{sec:relentropy}, the island formula implies that there exists substantial correlation between different fields in the Hawking radiation. In this appendix, we illustrate this point in more details.  Instead of dividing the Hawking radiation into spatial subsystems, we can also divide the Hawking radiation in a fixed region into subsets of fields. In other words, we are dividing the Hilbert space of some subregion into the tensor product of some smaller Hilbert spaces. This division is at least well defined when the Hawking radiation is in an asymptotic region, where we can neglect the dynamical gravity effects.

Let's consider a region $R$ that is big enough such that we have islands. Now we separate the fields in $R$ into a large portion (labeled by $L$) and a small portion (labeled by $S$). We take the number of degrees of freedom in $S$ to be much smaller than $L$. In the example of \S\ref{sec:example}, $S$ will be the massless fermion field under the $U(1)$ global symmetry, while $L$ will be the rest of the fields.

We could compute the entropies of the three density matrices $\rho_{\rm exact} (R_L \cup R_S)$, $\rho_{\rm exact}(R_L)$ and $\rho_{\rm exact} (R_S)$ using the QES prescription. For $\rho_{\rm exact} (R_L \cup R_S)$ and $\rho_{\rm exact}(R_L)$, we will find an island in the entanglement wedge:
\begin{equation}\label{SRLRS}
    S(\rho_{\rm exact} (R_L \cup R_S)) = \frac{A(\partial I)}{4G_N} + S (\rho_{\rm semi} (R_L \cup R_S \cup I_L \cup I_S)),
\end{equation}
\begin{equation}\label{SRL}
    S(\rho_{\rm exact} (R_L )) = \frac{A(\partial I)}{4G_N} + S (\rho_{\rm semi} (R_L \cup  I_L \cup I_S)),
\end{equation}
where we neglected the change in the position of the island due to the addition of the field $S$,
since it will only give subleading corrections. Importantly, in (\ref{SRL}), one gets contributions from both $I_L$ and $I_S$ in the island. This is because the island should be for all the fields, as it follows from the replica wormhole geometry which all the fields live on. (In other words, the emergent twist operators act on all fields.)
For $\rho_{\rm exact} (R_S)$, since its semi-classical entropy is small, there will not be an island:
\begin{equation}\label{SRS}
    S(\rho_{\rm exact} (R_S )) =  S (\rho_{\rm semi} (R_S)).
\end{equation}
Combining (\ref{SRLRS}) to (\ref{SRS}), one finds
\begin{equation}\label{Iexact}
    I_{\rm exact}( R_S : R_L   ) = I_{\rm semi} ( R_S : R_L \cup I_L \cup I_S).
\end{equation}
In particular, by the monotonicity of mutual information, one has
\begin{equation}
    I_{\rm exact}( R_S : R_L   ) > I_{\rm semi} ( R_S : R_L ),
\end{equation}
and the difference will be an order one amount. The difference comes from nonperturbative effects that are not present in the semiclassical description. Notice that in the vacuum state, the semiclassical mutual information $I_{\rm semi} ( R_S : R_L)$ will vanish in the limit that $S$ and $L$ are decoupled in the semiclassical theory. On the other hand, the right hand side of (\ref{Iexact}) reduces to $I_{\rm semi} ( R_S : I_S)$, 
which is nonzero and finite even when the coupling between $S$ and $L$ is negligible in the semiclassical description.

\section{Details on the relative entropy for massless free fermion in 2D}\label{app:modular}

As derived in \cite{Casini:2009vk,Arias:2018tmw}, for a two dimensional free massless fermion in the vacuum state, the modular Hamiltonian of the union of $I = (a_1,b_1)$ and $R= (a_2,b_2)$  takes the following form:
\begin{equation}\label{ModExpression}
\begin{aligned}
    K & = K_{\rm loc} + K_{\rm noloc},\\
    K_{\rm loc} & =  2 \pi \int_{I\cup R} \text{d} x\, \omega^{\prime}(x)^{-1} T(x), \\
    K_{\rm noloc} &  =  i 2 \pi \int_{I\cup R} \text{d} x\, \psi^{\dagger}(x)  \frac{\left(b_{1}-a_{1}\right)\left(a_{2}-b_{1}\right)\left(b_{2}-a_{1}\right)\left(b_{2}-a_{2}\right)}{\omega^{\prime}(x)^{2}\left(x-a_{1}\right)\left(x-a_{2}\right)\left(x-b_{1}\right)\left(x-b_{2}\right)} \\
    & \quad  \frac{1}{x\left(a_{1}+a_{2}-b_{1}-b_{2}\right)+\left(b_{1} b_{2}-a_{1} a_{2}\right)} \psi(\bar{x}),
\end{aligned}
\end{equation}
where $T(x) = \frac{1}{2} \left[ i \partial_x \psi^\dagger (x) \psi(x) - \psi^\dagger (x) i \partial_x \psi(x)\right]$ is the energy density operator, and $\omega(x), \bar{x}$ are given by
\begin{equation}
  \begin{aligned} \omega^{\prime}(x) &=\frac{1}{x-a_{1}}+\frac{1}{x-a_{2}}-\frac{1}{x-b_{1}}-\frac{1}{x-b_{2}} \\ \bar{x} &=\frac{a_{1} a_{2}\left(x-b_{1}-b_{2}\right)-b_{1} b_{2}\left(x-a_{1}-a_{2}\right)}{x\left(a_{1}+a_{2}-b_{1}-b_{2}\right)+\left(b_{1} b_{2}-a_{1} a_{2}\right)} \end{aligned}.
\end{equation}
The formulas are the same for $\psi_+$ and $\psi_-$, so we've omitted the chirality of the fermion. As discussed in (\ref{relModular}), we are interested in computing the expectation value $\bra{0} K_{\rm noloc} \ket{0}$. Using
\begin{equation}
    \bra{0} \psi^\dagger (x) \psi (y) \ket{0} = \frac{i}{2\pi} \frac{1}{x-y},
\end{equation}
when $x\neq y$, we can compute the integral in (\ref{ModExpression}), and find
\begin{equation}\label{Kexpectation}
    \bra{0} K_{\rm noloc} \ket{0} = -\frac{1}{4} \left( 1 + \frac{ (2\eta - 1)\arctan \sqrt{\frac{\eta}{1-\eta}} }{ \sqrt{\eta (1-\eta)} } \right),
\end{equation}
where the cross ratio $\eta$ is defined as
\begin{equation}
    \eta \equiv \frac{(b_1 - a_1) (b_2 - a_2)}{(a_2-a_1) (b_2 -b_1)}. 
\end{equation}
When computing the integral, it is convenient to set $\{a_1,b_1,a_2,b_2\} = \{0,\eta,1,\infty\}$. (\ref{Kexpectation}) is the contribution from one of the chiral modes, so to get the total contribution, we multiply it by two. 

\section{Global symmetry in JT gravity + EoW branes \label{app:westcoast} }
In this section, we explore issues related to global symmetry violations in the ``West coast'' model \cite{Penington:2019kki}, namely JT gravity with end-of-the-world (EoW) branes. We refer the reader to \cite{Penington:2019kki} for definitions and conventions. Some similar comments could be made in the simplified model of \cite{Marolf:2020xie}.

\subsection{coarse-grained entropy}
A natural question is whether one can define a coarse-grained density matrix $\tilde{\rho}_R$ of the radiation, whose entropy does not follow the Page curve but instead is given by the naive, non-minimal quantum extremal surface.
In the setup of \cite{Penington:2019kki}, we want the coarse-grained entropy to be $\approx \log k$, even when $k$ is bigger than $e^{S_0}$.
Here we point out that if we have a large symmetry group, we can use group-averaging to define a coarse-grained density matrix:
\eqn{\tilde{\rho}_R = \int dg \, U(g) \rho_R U({g\inv} ) .}
Note that the group-averaging is a completely positive trace preserving (CPTP) map. It takes any density matrix into a density matrix which is invariant under the symmetry group. 
The CPTP property ensures that
\eqn{S(\tilde{\rho}_R) \ge S(\rho_R).}

For a simple model where we can explicitly compute $S(\tilde{\rho}_R)$, consider the west coast model \cite{Penington:2019kki}. From the bulk (semi-classical) perspective, the model is defined by JT gravity +  end-of-the-world (EoW) branes. 
The EoW branes carry an index $i$ such that \eqn{\ev{i|j}_\text{bulk} = \delta_{ij}.} This equation should not be confused with a statement about the actual boundary states that correspond to these branes, which are not exactly orthonormal. The content of this equation is simply to prescribe rules for evaluating the bulk path integral. The point we would like to make here is that the bulk theory at the semi-classical level has a $U(k)$ global symmetry, where the EoW branes transform in the fundamental.

Now let us consider the replica wormhole computation of $S(\tilde{\rho}_R)$. We will start by computing the R\'{e}nyi entropies. 
Let us consider the 2-replica computation of $\tr (\tilde{\rho}_R^2)$. 
Since the overall state of the black hole and the radiation is an entangled state \eqn{\ket{\Psi}={1 \over \sqrt{k}} \sum_{i=1}^k \ket{\psi_i}_\textrm{B} \ket{i}_\textrm{R},}
acting with the symmetry generator on the radiation $\ket{i} \to U \ket{i}_\mathrm{R} = U_{ji} \ket{j}_\mathrm{R}$ is equivalent to acting on the black hole states $\ket{\psi_i}_\textrm{B} \to \sum_j U_{ji} \ket{\psi_j}_\textrm{B}$.
Clearly if we consider the disconnected saddle, the symmetry generators act trivially. But for the connected saddle, we actually get the constraint $i = j$. So whereas
\eqn{\operatorname{Tr}\left(\rho_{\mathrm{R}}^{2}\right)=\frac{k Z_{1}^{2}+k^{2} Z_{2}}{\left(k Z_{1}\right)^{2}}=\frac{1}{k}+\frac{Z_{2}}{Z_{1}^{2}},\\
\operatorname{Tr}\left( \tilde{\rho}_{\mathrm{R}}^{2}\right)=\frac{k Z_{1}^{2}+k Z_{2}}{\left(k Z_{1}\right)^{2}}=\frac{1}{k}\lp 1 +\frac{Z_{2}}{Z_{1}^{2}}\rp .\\
}
Notice that the connected saddle never dominates, no matter how big $k$ is.

In fact, from the boundary point of view we can compute the exact density matrix after group averaging:
\eqn{\rho_{\mathrm{R}} &= \frac{1}{k} \sum_{i, j=1}^{k}\ket{j} \bra{i}_{\mathrm{R}} \ev{\psi_i|\psi_j}_{\mathrm{B}}.\\
\tilde{\rho}_{\mathrm{R}}&=\frac{1}{k} \sum_{i,j=1}^{k} \int dU  \, U\ket{j} \bra{i}_{\mathrm{R}} U^\dagger \ev{\psi_i|\psi_j}_{\mathrm{B}}\\
&= {1 \over k} \sum_{i} \ket{i}\bra{i}_R\ev{\psi_i|\psi_i}\\
&\approx {1 \over k} \sum_{i} \ket{i}\bra{i}_R.
}
Note that this is almost a completely mixed density matrix. (It is not perfectly mixed because the norms of the states  $\ev{\psi_i|\psi_i}$ have small fluctuations. However, to leading order in $e^{-S_0}$, this effect is negligible.) 
Notice that at large $k$, this gives a concrete meaning to a sub-dominant QES.

\subsection{Relative entropy}
Let us start with computing the relative purity $\tr U \rho U^\dagger 
\rho$ in the West coast model. Up to a normalization,
\eqn{\tr U \rho U^\dagger 
\rho &= U_{ij} \rho_{jk} U^\dagger_{kl} \rho_{li} = \\
&= U_{ij} U_{kl}^\dagger \ev{\psi_j| \psi_k} \ev{\psi_l| \psi_i}\\
&= \sum_{i,k} \delta_{ik} Z_1^2 + U_{ii}U_{kk}^\dagger Z_2
}
Notice that the unitary does not affect the disconnected contribution. Furthermore, when we set the group element to the identity $U = I$, we recover the result in \cite{Penington:2019kki}.

Now we can also use the QES prescription to compute the relative entropy.
At small $k$, there is no island and the QES prescription gives 0. 
Since the semi-classical density matrix is a pure state:
\eqn{\rho_\text{semi} = \sum_{i,j}( \ket{\psi_i}\bra{\psi_j}  \otimes \ket{i} \bra{j} )  }
the relative entropy would be infinite if $U\neq 1$:
\eqn{S(U_R \rho_\textrm{semi}(R \cup I) U^\dagger_R|\rho_\textrm{semi}(R \cup I) )= \infty, \quad U \ne 1.}
In the exact description, the density matrix of the radiation is not pure. We therefore expect a finite answer, so the relative entropy formula is not immediately applicable. It would be interesting to apply the planar resummation techniques of \cite{Penington:2019kki} to compute the finite answer.


\subsection{Global symmetry in the ensemble}\la{app:C3}
Here we would like to make some distinctions about the various kinds of symmetries in JT gravity $+$ EoW branes. 
\begin{enumerate}
    \item The simplest symmetry is the semi-classical symmetry of the theory. This means that if we only include the trivial, disconnected topologies in the path integral, we will have a matter theory which has an exact global $U(k)$ symmetry. In particular, the overlaps between different brane states $\ev{\psi_i| \psi_j} = \delta_{ij}$ are exactly invariant under $\ket{\psi_i} \to \sum_j U_{ij} \ket{\psi_j}$.
    
    \item If we include higher topologies in the path integral, it has been shown that the path integral is dual to an ensemble of theories. The ensemble is a probability distribution over the Hamiltonian (a random matrix) and a set of pure states ($k$ random vectors). If we consider one instance of the disorder average, the resulting theory will generically not have a $U(k)$ global symmetry. For example, the inner products between different pure states will be slightly different.
    \item Nevertheless, the ensemble has a $U(k)$ symmetry. Note that unlike in point 1, the $U(k)$ symmetry acts in {\it theory space}. It is not the symmetry group of an individual theory. Specifying each theory means specifying the values of the Hamiltonian and the pure state. The statement about the symmetry of the ensemble is that if we take such one instance of the ensemble $t_1$, and then act with $g \in U(k)$ on the random vectors to obtain a new theory $t_2 =g (t_1)$, the probability $P(t_2) = P(t_1)$. If we think of the ensemble as a third-quantization \cite{Marolf:2020xie}, we might say that $U(k)$ is a symmetry in the third-quantized theory.
 \end{enumerate}

\bibliographystyle{JHEP}
\bibliography{cite}

\end{document}